\documentclass[10pt, pre, twocolumn, showpacs, aps]{revtex4-1}
\usepackage{amsmath,amssymb}
\usepackage[dvipdfmx]{graphicx,color}
\usepackage{ulem}
\usepackage{pgf,tikz}
\usepackage{here}
\usepackage{comment}

\newcommand{\diff}{\mathrm{d}}
\newcommand{\ave}[1]{\langle #1 \rangle}
\newcommand{\aveave}[1]{\langle\hspace*{-0.2em}\langle #1 \rangle\hspace*{-0.2em}\rangle}

\begin{document}
\title
  {Classification of bifurcation diagrams
  in coupled phase-oscillator models
  with asymmetric natural frequency distributions}
\author{Ryosuke Yoneda and Yoshiyuki Y. Yamaguchi}
\affiliation{Graduate School of Informatics, Kyoto University, 606-8501 Kyoto, Japan}

\begin{abstract}
  Synchronization among rhythmic elements is modeled
  by coupled phase-oscillators
  each of which has the so-called natural frequency.
  A symmetric natural frequency distribution induces
  a continuous or discontinuous synchronization transition
  from the nonsynchronized state, for instance.
  It has been numerically reported that asymmetry in the natural frequency
  distribution brings new types of bifurcation diagram
  having, in the order parameter,
  oscillation or a discontinuous jump
  which emerges from a partially synchronized state.
  We propose a theoretical classification method of five types
  of bifurcation diagrams including the new ones,
  paying attention to the generality of the theory.
  The oscillation and the jump from partially synchronized states
  are discussed respectively
  by the linear analysis around the nonsynchronized state
  and by extending the amplitude equation
  up to the third leading term.
  The theoretical classification is examined by comparing
  with numerically obtained one.
\end{abstract}
\maketitle

\section{Introduction}



Synchronization among rhythmic elements is observed in various fields of nature:
metronomes \cite{pantaleone2002},
flashing fireflies \cite{smith1935,buck1968},
frog choruses \cite{aihara2014},
and Josephson junction arrays \cite{wiesenfeld1996,wiesenfeld1998}.
The synchronization has been theoretically studied
through coupled phase-oscillator models \cite{winfree1967,kuramoto1975}.
The Kuramoto model \cite{kuramoto1975}, a paradigmatic model,
describes the bifurcation from the nonsynchronized state
to partially synchronized states
and the type of bifurcation depends on the distribution of
the so-called natural frequencies each of which is assigned
for each phase-oscillator.
The bifurcation is continuous
if the natural frequency distribution is symmetric and unimodal
\cite{kuramoto1975,strogatz2000,chiba2015},
but the continuity breaks if the distribution is
flat at the peak \cite{basnarkov-urumov-07,bastian-18}
or bimodal \cite{martens2009}.
A bimodal distribution also yields temporally oscillating states
\cite{martens2009}.

A significant part of previous studies assumes symmetry
of natural frequency distributions,
and bifurcation
is found as the synchronization transition
  referring to the transition
  from the nonsynchronized state to partially synchronized states.
  Interestingly, symmetry breaking brings new types of bifurcations
  that emerge not from the nonsynchronized state
  but from partially synchronized states
  and are followed by a discontinuous jump or oscillation
  of the order parameter \cite{terada2017}.
The new types of bifurcations have been observed
by performing numerical simulations,
and this paper aims to propose a theoretical explanation
of the new types of bifurcations.
For analyzing bifurcations,
we have three theoretical methods
based on the large population limit
and the equation of continuity.



The first method is the self-consistent equation for the order parameter
of system \cite{kuramoto1975,strogatz2000,pazo-05,basnarkov-urumov-07,park-kahng-19}.
A stationary solution to the equation of continuity
is formally written by using an unknown value of the order parameter
and the value is determined self-consistently.
The self-consistent strategy is powerful to obtain
stationary values of the order parameter,
but the oscillating states are not captured.
Moreover, writing down stationary states is not straightforward
when a coupling function between a pair of oscillators
has several sinusoidal functions
\cite{komarov-pikovsky-13,komarov-pikovsky-14}.


The second method is the Ott-Antonsen ansatz \cite{oa2008,oa2009}.
  This ansatz reduces the equation of continuity
  to a simpler partial differential equations.
  This strategy is applicable for the Kuramoto model
  and its variances \cite{lu-etal-16,akao-19},
  and is extremely useful when we use a single Lorentzian
  as the natural frequency distribution,
  thereby resulting a real two-dimensional reduced system.
The dimension of the reduced system becomes higher
as the distribution becomes complicated
\cite{martens2009,terada2017,bastian-18},
and searching for stationary and oscillating states becomes difficult
accordingly.


The third method is the amplitude equation,
which describes dynamics projected onto the unstable manifold
of the nonsynchronized state \cite{crawford1994}.
In contrast to the previous two methods,
the amplitude equation is widely useful
in the Kuramoto model \cite{crawford1994},
in a coupled phase-oscillator model with a generic coupling function
\cite{crawford1995},
in the time-delayed Kuramoto model \cite{metivier-19},
in an extended Kuramoto model with inertia \cite{barre2016},
and in Hamiltonian systems
\cite{crawford1994b,crawford1995b,barre-metivier-yamaguchi-16}.
Other advantages of the amplitude equation are
that one-dimensional dynamics is sufficient
to judge the continuity of bifurcation
and that any natural frequency distributions are equally tractable.

The first and second methods profit from a single sinusoidal coupling function,
but such a simple coupling function is not always the case
\cite{hansel-mato-meunier-93,crawford1995,kiss-zhai-hudson-05,kiss-zhai-hudson-06}.
We, therefore, adopt the third method, the amplitude equation,
to explain the new types of bifurcations.
One basic block of the amplitude equation is the linear analysis
around the nonsynchronized state,
and the linear analysis gives, for instance,
eigenvalues.
We will demonstrate that the linear analysis is also useful
to explain a mechanism for the emergence of an oscillating state.


The method proposed in this paper is in principle applicable to general systems
beyond the Kuramoto model and its variances,
but we investigate the Kuramoto model 
to illustrate the usefulness of the method.
One reason is that the new types of bifurcations are reported
in the Kuramoto model, and this choice makes it straightforward
to compare the theory with the previous work
\cite{terada2017}.
Another reason is that, as we discussed above,
the reduction by the Ott-Antonsen ansatz
permits us to obtain accurate classification,
which helps to examine the theory.

This paper is organized as follows.
In Sec. \ref{sec:model}, we introduce the Kuramoto model
and its large population limit
written by the equation of continuity.
The considering family of the natural frequency distribution,
which is characterized by two parameters, is also exhibited.
In Sec. \ref{sec:numerics}, we numerically divide the parameter plane 
into five domains corresponding to types of bifurcation diagrams
and prepare the reference parameter plane to examine the theory.
This numerical search is performed by using the Ott-Antonsen reduction.
We stress that this reduction is used only for obtaining
the reference parameter plane and is not used in our theory.
In Sec. \ref{sec:analysis}, linear and nonlinear analyses
of the equation of continuity are shortly reviewed.
With the aid of these analyses,
we propose ideas to identify the domains on the parameter plane
and report the theoretical consequence in Sec. \ref{sec:criteria}.
Finally, we summarize this paper in Sec. \ref{sec:conclusion}.

\section{Model}
\label{sec:model}
The Kuramoto model is
expressed by the $N$-dimensional ordinary differential equation,
\begin{align}
  \frac{\diff\theta_{j}}{\diff t}
  = \omega_{j} - \frac{K}{N}\sum_{k=1}^{N}\sin\left(\theta_{j}-\theta_{k}\right),
  \quad
  (j=1,\cdots,N).
  \label{eq:kuramoto}
\end{align}
The real constant $K>0$ is the coupling constant,
$\theta_{j}\in (-\pi,\pi]=\mathbb{S}^{1}$ and $\omega_{j}\in\mathbb{R}$
are respectively the phase and the natural frequency of the $j$th oscillator.
The natural frequencies $\{\omega_{j}\}$ obey
a probability distribution function $g(\omega)$.
  By using the parameters $\gamma_{1},\gamma_{2}>0$ and $\Omega\geq 0$,
we introduce a family of $g(\omega)$ as
\begin{align}
  \label{eq:g}
  g(\omega)
  = \dfrac{C}{[(\omega-\Omega)^{2}+\gamma_{1}^{2}][(\omega+\Omega)^{2}+\gamma_{2}^{2}]}
\end{align}
to systematically consider unimodal and bimodal,
and symmetric and asymmetric distributions \cite{terada2017,terada2018}.
The family of $g(\omega)$ consists of rational functions,
which is useful to draw the reference parameter plane by using 
the Ott-Antonsen ansatz, but is not crucial in our methodology.
The normalization constant
\begin{align}
  C=\frac{\gamma_{1}\gamma_{2}[(\gamma_{1}+\gamma_{2})^2+4\Omega^2]}{\pi(\gamma_{1}+\gamma_{2})}
\end{align}
is determined from the condition
\begin{align}
  \int_{-\infty}^{\infty}g(\omega) \diff\omega = 1.
\end{align}
The three parameters $\gamma_{1},\gamma_{2},$ and $\Omega$
are assumed to be positive.
By scaling the variables $t,\omega_{j},K,$ and $\gamma_{1}$,
we may set $\gamma_{2}=1$ without loss of generality,
and the family of $g(\omega)$ is characterized
by a point on the parameter plane $(\gamma_{1},\Omega)$.
  The parameter plane is further restricted
  into the region of $\gamma_{1}\leq 1$
  by making the one-to-one correspondence between
  the regions of $\gamma_{1}>1$ and $\gamma_{1}<1$ as follows.
  The equation of motion, \eqref{eq:kuramoto},
  is invariant under the transformations of $\theta_{j}\to - \theta_{j}$
  and $\omega_{j} \to - \omega_{j}$,
  and the latter induces the transformation of $g(\omega)\to g(-\omega)$.
  The equation of motion is also invariant
  under the change of time scale $t\to t/\tau$ with
  $\omega_{j} \to \tau\omega_{j}$ and $K \to \tau K$.
  These changes do not modify type of the bifurcation diagram for $\tau>0$
  and the transformed natural frequency distribution is
  \begin{equation}
    g(-\omega) \propto \dfrac{1/\tau^{4}}{[(\omega+\Omega/\tau)^{2}+(\gamma_{1}/\tau)^{2}][(\omega-\Omega/\tau)^{2}+(\gamma_{2}/\tau)^{2}]}.
  \end{equation}
  The point $(\gamma_{1},\Omega)$, therefore, corresponds to
  the point $(1/\gamma_{1},\Omega/\gamma_{1})$
  by selecting $\tau=\gamma_{1}$ under $\gamma_{2}=1$.
The line $\gamma_{1}=1$ gives a family of symmetric distributions.

The complex order parameter $z$ is defined by
\begin{align}
  z=re^{i\psi}=\frac{1}{N}\sum_{j=1}^{N}e^{i\theta_{j}},
  \quad
  (r,\psi\in\mathbb{R}).
  \label{eq:order}
\end{align}
The absolute value $r$ of $z$ measures the extent of synchronization.
If all the oscillators distribute uniformly on $\mathbb{S}^{1}$,
then $r\simeq 0$.
If $r\simeq 1$, the majority of oscillators
gathers around a point on $\mathbb{S}^{1}$.

In the limit of large population $N\to\infty$,
by the conservation of the number of oscillators,
the equation of motion \eqref{eq:kuramoto}
can be written in the equation of continuity \cite{lancellotti2004},
\begin{align}
\begin{aligned}
&\frac{\partial F}{\partial t}+\frac{\partial}{\partial\theta}(v[F]F)=0,\\
    &v[F]
    = \omega - K \int_{-\infty}^{\infty} \diff\omega' \int_{-\pi}^{\pi} \diff\theta'~
    \sin(\theta-\theta')F(\theta',\omega',t),
\end{aligned}
  \label{eq:kuramoto-inf}
\end{align}
where $F(\theta,\omega,t)$ is the probability distribution function
of $\theta$ and $\omega$ at the time $t$.
In other words, $F(\theta,\omega,t)d\theta ~\diff\omega$
represents the fraction
of oscillators having phases between $\theta$ and $\theta+d\theta$
and natural frequencies between $\omega$ and $\omega+~\diff\omega$
at the time $t$.
From the normalization condition $\int Fd\theta ~\diff\omega=1$, we have
\begin{align}
  \int_{-\pi}^{\pi} F(\theta,\omega,t)\diff\theta=g(\omega).
\end{align}
In this limit the order parameter is expressed by
\begin{align}
  z=re^{i\psi}=\int_{-\infty}^{\infty} \diff\omega \int_{-\pi}^{\pi} \diff\theta
  ~ F(\theta,\omega,t) e^{i\theta}.
\end{align}

\section{Numerical Classification on Parameter plane}
\label{sec:numerics}

The aim of this section is to classify numerically the parameter plane
$(\gamma_{1},\Omega)$ into five types of bifurcation diagrams
\cite{terada2018} before developing the theory.
Direct $N$-body simulations of the model \eqref{eq:kuramoto}
include finite-size fluctuations, which make it difficult
to judge the types of bifurcation diagrams.
For eliminating the finite-size fluctuations,
we perform the Ott-Antonsen reduction \cite{oa2008,oa2009},
which reduces the equation of continuity
to a real four-dimensional differential equation
for the family of $g(\omega)$, \eqref{eq:g}.



\subsection{Ott-Antonsen reduction}
\label{oa}

The reduced equation by using the Ott-Antonsen ansatz is expressed as
\cite{terada2017}
\begin{align}
\begin{aligned}
    &\frac{\diff z_{1}}{\diff t}=(i\Omega-\gamma_{1})z_{1}-\frac{K}{2}(z^{*}z_{1}^{2}-z),\\
    &\frac{\diff z_{2}}{\diff t}=-(i\Omega+\gamma_{2})z_{2}-\frac{K}{2}(z^{*}z_{2}^{2}-z).
\end{aligned}
  \label{eq:oa-12}
\end{align}
The two variables $z_{1}$ and $z_{2}$ are complex
and the complex order parameter $z$ is written as
\begin{align}
  z=k_{1}z_{1}+k_{2}z_{2}
\end{align}
with the complex constants $k_{1}$ and $k_{2}$ defined by
\begin{align}
\begin{aligned}
&k_{1}=\frac{\gamma_{2}[2\Omega-i(\gamma_{1}+\gamma_{2})]}{(\gamma_{1}+\gamma_{2})[2\Omega+i(\gamma_{1}-\gamma_{2})]},\\
    &k_{2}=\frac{\gamma_{1}[2\Omega+i(\gamma_{1}+\gamma_{2})]}{(\gamma_{1}+\gamma_{2})[2\Omega+i(\gamma_{1}-\gamma_{2})]}.
\end{aligned}
  \label{eq:k1k2}
\end{align}
$z^{\ast}$ represents the complex conjugate of $z$.
In the later computations we set $\gamma_{2}=1$ without loss of generality
as we mentioned in the previous section \ref{sec:model},
but we kept $\gamma_{2}$ free to show the dependence explicitly.
See Appendix \ref{sec:oa} for details of the reduction.

\subsection{Bifurcation diagrams}

Numerical integration of the reduced system \eqref{eq:oa-12}
is performed by using the fourth-order Runge-Kutta algorithm
with the time step $\Delta t=0.01$.
For a given set of $(\gamma_{1},\Omega)$,
we start from $K=0$ and increase the value up to $K=10$
with the step $\Delta K=0.0001$.
This increasing process is called the forward process.
At each $K$, the time average and standard deviation of the order parameter
are taken in the time interval $t\in [4500,5000]$ to avoid a transient regime.
The final state at $t=5000$ is used
as the initial state at the successive value of $K$.
If the final state is the origin,
then we shift the initial state from the origin to $z_{1}=z_{2}=0.01$
to escape from the trivial stationary state.
After arriving $K=10$, we decrease the value of $K$
from $K=10$ to $K=0$ to check existence of a hysteresis,
which reveals a discontinuous bifurcation.
This decreasing process is called the backward process.

\begin{figure}
  \centering
  \includegraphics[width=8cm]{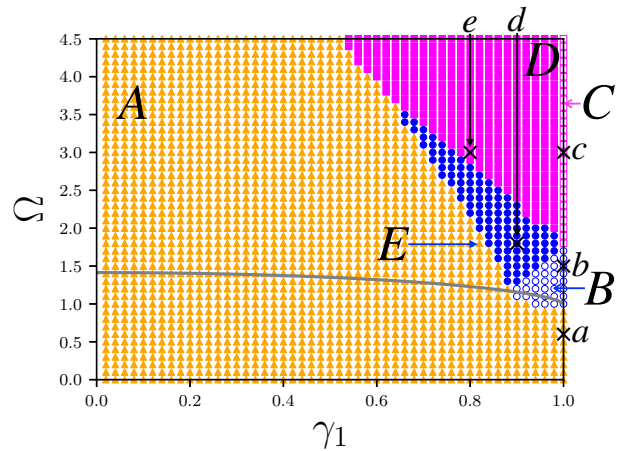}  
  \caption{Classification of the parameter plane
    ($\gamma_{1},\Omega$) into the five domains:
    A (filled orange triangles), B (open blue circles),
    C (open magenta squares),
    D (filled magenta squares), and E (filled blue circles).
    This classification is obtained by performing numerical simulations
    of the reduced system \eqref{eq:oa-12}.
    The gray line represents the borderline between
    the unimodal and bimodal natural frequency distributions $g(\omega)$.
    The bifurcation diagrams at the five points marked by the crosses
    are reported in Fig.~\ref{fig:bif-diagrams}.
    }
  \label{fig:phase-diagram}
\end{figure}

\begin{figure}
  \centering
  \includegraphics[width=8cm]{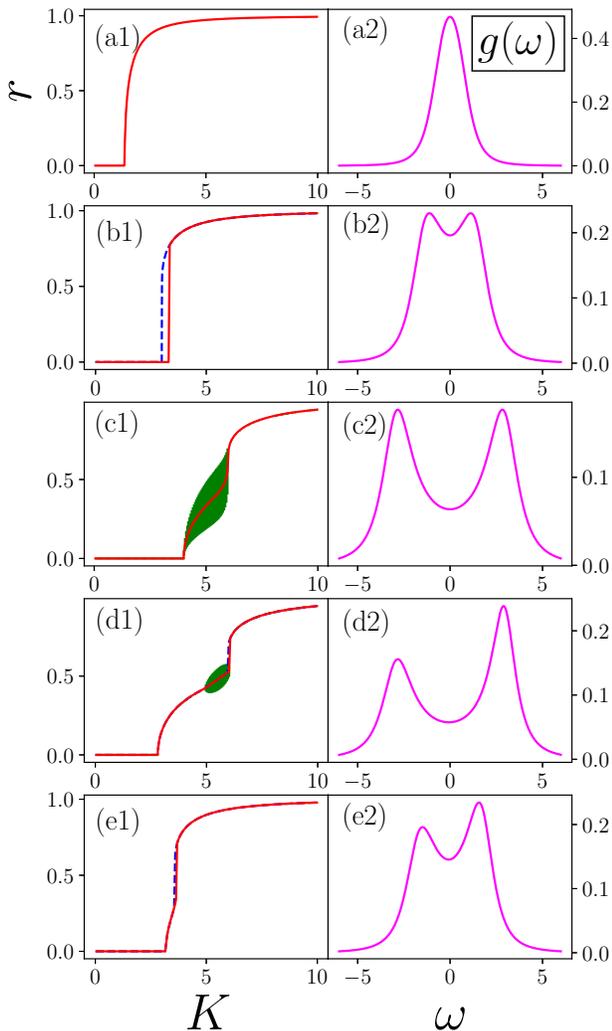}
  \caption{Five bifurcation diagrams at the five points
    marked on the parameter plane in Fig.~\ref{fig:phase-diagram}
    and the corresponding natural frequency distributions $g(\omega)$.
    The values of parameters $(\gamma_{1},\Omega)$ are
    (a) $(1.0,0.6)$, (b) $(1.0,1.5)$, (c) $(1.0,3.0)$,
    (d) $(0.8,3.0)$, and (e) $(0.9,1.8)$.
    In the panels (a1),(b1),(c1),(d1), and (e1),
    the red solid line and the blue dashed line
    respectively denote the forward
    and the backward process.
    In the panels (a1) and (c1), the backward process line
    collapses with the forward process line.
    Standard deviations of the order parameter
    are represented by the green vertical bars,
    but they are not visible in (a1), (b1), and (e1).
    The panels (a2),(b2),(c2),(d2), and (e2)
    show the natural frequency distributions
    $g(\omega)$ against $\omega$.
    }
  \label{fig:bif-diagrams}
\end{figure}

The parameter plane $(\gamma_{1},\Omega)$ is classified
into five domains as shown in Fig.~\ref{fig:phase-diagram}
\cite{terada2018},
which domains correspond to the five types of bifurcation diagrams
reported in Fig.~\ref{fig:bif-diagrams}.
The symmetry line $\gamma_{1}=1$,
in particular,
is classified into the three intervals included in the domains A, B, and C.
The separating points $\Omega=1$ and $\sqrt{3}$
are obtained by applying the amplitude equation
and the eigenvalue analysis respectively, which are explained later. 
The asymmetry region $\gamma_{1}<1$
includes the new domains D and E
with two known domains A and B.
The goal of this paper is to reproduce
the parameter plane theoretically
for understanding mechanisms yielding the new domains D and E.

\section{Theoretical analyses of equation of continuity}
\label{sec:analysis}

We shortly review linear and nonlinear analyses
of the equation of continuity \eqref{eq:kuramoto-inf}
around the nonsynchronized state $f^{0}$, which is explicitly written as
\begin{align}
  f^{0}(\omega) = \frac{g(\omega)}{2\pi}.
\end{align}
It is straightforward to check stationarity of $f^{0}$ as
\begin{align}
  \frac{\partial}{\partial\theta}(v[f^{0}]f^{0}) = 0
\end{align}
from the fact $v[f^{0}] = \omega$.
We expand the equation of continuity by substituting
\begin{align}
  F(\theta,\omega,t) = f^{0}(\omega) + f(\theta,\omega,t). 
\end{align}
The perturbation $f$ is governed by the equation
\begin{align}
  \frac{\partial f}{\partial t}=\mathcal{L}f+\mathcal{N}[f],
  \label{eq:expansion-f}
\end{align}
where the linear part is
\begin{align}
  \label{eq:linear-L}
  \mathcal{L}f
  = - \omega \dfrac{\partial f}{\partial\theta}
  - K f^{0} \dfrac{\partial}{\partial\theta}
  {\rm Im} \left( z(t) e^{-i\theta} \right)
\end{align}
and the nonlinear part is
\begin{align}
  \label{eq:nonlinear-N}
  \mathcal{N}[f]
  = - K \dfrac{\partial}{\partial\theta} \left[
    f~ {\rm Im}\left( z(t) e^{-i\theta} \right) \right]
\end{align}
with the order parameter
\begin{align}
  z(t) = \int_{-\infty}^{\infty} \diff\omega \int_{-\pi}^{\pi} \diff\theta
  e^{i\theta} f(\theta,\omega,t).
\end{align}
Note that the equation \eqref{eq:expansion-f} is an exact transform
from the equation of continuity \eqref{eq:kuramoto-inf}.
The linear part will be used to explain the oscillating state
emerging from a partially synchronized state
as well as a basic block of the amplitude equation.

\subsection{Linear analysis}
\label{eq:linear-eigenvalues}

The perturbation $f$ is a periodic function of $\theta$
and is expanded into the Fourier series as
\begin{align}
  f(\theta,\omega,t)
  = \sum_{k\in\mathbb{Z}} \widetilde{f}_{k}(\omega,t) e^{ik\theta}.
  \label{eq:f-Fourier}
\end{align}
The linear analysis of \eqref{eq:linear-L}
can be performed independently in each Fourier mode $k$
because the nonsynchronized state $f^{0}(\omega)$ does not depend on $\theta$.
The Fourier modes $k\neq \pm 1$ give only rotations,
and instability comes from the modes $k=\pm 1$.
The eigenvalues for the modes $k=\pm 1$ are obtained
as roots of the spectral functions
\begin{align}
  \Lambda_{\pm 1}(\lambda)
  = 1 - \frac{K}{2} \int_{-\infty}^{\infty}
  \frac{g(\omega)}{\lambda\pm i\omega} \diff\omega.
  \label{eq:spectral-function}
\end{align}
See Appendix \ref{sec:spectral-functions} for derivations.
If the real part of a root is positive, the eigenvalue induces instability.
We call such an eigenvalue as an unstable eigenvalue,
which is the target of the amplitude equation introduced
in the next subsection \ref{sec:amplitude-equation}.

We define the synchronization transition point $K_{\rm c}$
at which the nonsynchronized state $f^{0}$ changes stability.
This definition suggests that the eigenvalue having the largest real part
must be on the imaginary axis at $K_{\rm c}$.
A pure imaginary eigenvalue, however, induces a singularity
in the integrands of the spectral functions.
To avoid the singularity, we perform the analytic continuation
of $\Lambda_{\pm 1}(\lambda)$
and denote the continued functions as $D_{\pm 1}(\lambda)$.
See Appendix \ref{sec:analytic-continuation} for the continuation.

We give three remarks.
First, the continuation does not modify the spectral functions
in the region ${\rm Re}(\lambda)>0$,
and a root of $D_{\pm 1}(\lambda)$ whose real part is positive
is also a root of $\Lambda_{\pm 1}(\lambda)$.
Second, a root of $D_{\pm 1}(\lambda)$, however, may not be an eigenvalue
if it is on the region ${\rm Re}(\lambda)\leq 0$,
and the root is called a fake eigenvalue accordingly.
Third, the relations
\begin{align}
  \Lambda_{-1}(\lambda^{\ast}) = \Lambda_{1}^{\ast}(\lambda)
  \label{eq:relation-Lambda}
\end{align}
and 
\begin{align}
  D_{-1}(\lambda^{\ast}) = D_{1}^{\ast}(\lambda)
  \label{eq:relation-D}
\end{align}
hold, where $\Lambda_{1}^{\ast}(\lambda)$ is, for instance,
the complex conjugate of $\Lambda_{1}(\lambda)$.
These relations imply that $\lambda^{\ast}$ is a (fake) eigenvalue
if $\lambda$ is.

For the family of natural frequency distributions \eqref{eq:g},
the equation $D_{1}(\lambda)=0$ leads a quadratic equation
of $\lambda$ and the two roots are denoted by $\lambda_{1}$ and $\lambda_{2}$
ordered as ${\rm Re}(\lambda_{1})\geq {\rm Re}(\lambda_{2})$.
In this paper, the two (fake) eigenvalues $\lambda_{1}$ and $\lambda_{2}$
are called the first and the second (fake) eigenvalues respectively.
See Appendix~\ref{sec:Kc} for the quadratic equation of $\lambda$,
the determination of the critical point $K_{\rm c}$,
and the number of unstable eigenvalues.

\subsection{Amplitude equation}
\label{sec:amplitude-equation}

We assume that the linear operator $\mathcal{L}$
has only one pair of unstable eigenvalues,
$\lambda_{1}$ and $\lambda_{1}^{\ast}$
coming from the relation \eqref{eq:relation-D},
and that the corresponding eigenfunctions are
$\Psi(\theta,\omega)$ and $\Psi^{\ast}(\theta,\omega)$ respectively.
To derive the amplitude equation, we expand $f$ into
\begin{align}
  f(\theta,\omega,t) = A(t) \Psi(\theta,\omega)
  + A^{\ast}(t) \Psi^{\ast}(\theta,\omega) + H(\theta,\omega, A, A^{\ast}).
  \label{eq:expand-f}
\end{align}
The amplitude $A$ relates to the order parameter $z$ as
\begin{align}
  z = 2\pi A^{\ast} + O(|A|^{3}),
\end{align}
and the asymptotic value of $z$ is approximately obtained by considering
temporal evolution of the amplitude $A(t)$.
The function $H$ represents the unstable manifold of
the nonsynchronized state $f^{0}$.
In other words, $H$ represents the height of the unstable manifold
from the eigenspace ${\rm Span}\{\Psi,\Psi^{\ast}\}$.
We assume that the unstable manifold is tangent
to the eigenspace ${\rm Span}\{\Psi,\Psi^{\ast}\}$
at $A=A^{\ast}=0$ and $H=O(|A|^{2})$ accordingly.

For deriving the amplitude equation,
we introduce the adjoint linear operator $\mathcal{L}^{\dagger}$ of $\mathcal{L}$
defined by
\begin{align}
  \left( \mathcal{L}^{\dagger} f_{1}, f_{2} \right)
  = \left( f_{1}, \mathcal{L} f_{2} \right),
  \quad
  \mathrm{ for }~\forall f_{1},f_{2},
\end{align}
where the inner product $(\cdot,\cdot)$ is defined by
\begin{align}
  \left( f_{1}, f_{2} \right)
  = \int_{-\infty}^{\infty} \diff\omega \int_{-\pi}^{\pi} \diff\theta
  f_{1}^{\ast}(\theta,\omega) ~ f_{2}(\theta,\omega).
\end{align}
Let $\widetilde{\Psi}$ and $\widetilde{\Psi}^{\ast}$
be the eigenfunctions of $\mathcal{L}^{\dagger}$
corresponding to the eigenvalues
$\lambda_{1}^{\ast}$ and $\lambda_{1}$ respectively.
We can choose the eigenfunctions satisfying the relations
\begin{align}
  \label{eq:normalization-tildePsi}
  \begin{aligned}
    & \left( \widetilde{\Psi}, \Psi \right) = 1,
     \quad
     \left( \widetilde{\Psi}, \Psi^{\ast} \right) = 0, \\
     & \left( \widetilde{\Psi}^{\ast}, \Psi \right) = 0,
     \quad
     \left( \widetilde{\Psi}^{\ast}, \Psi^{\ast} \right) = 1,
  \end{aligned}
\end{align}
without loss of generality. We also have
\begin{align}
  \left( \widetilde{\Psi}, H \right)
  = \left( \widetilde{\Psi}^{\ast}, H \right)
  = 0
  \label{eq:orthogonality-Psi-H}
\end{align}
because $H$ does not belong to the eigenspace ${\rm Span}\{\Psi,\Psi^{\ast}\}$.
See Appendix \ref{sec:adjoint} for the explicit expressions
of the adjoint operator and the eigenfunctions.

Substituting the expansion \eqref{eq:expand-f}
into the equation of continuity \eqref{eq:kuramoto-inf}
and using the relations \eqref{eq:normalization-tildePsi}
and \eqref{eq:orthogonality-Psi-H},
we have the amplitude equation as
\begin{align}
  \label{eq:amplitude-equation-general}
  \frac{{\rm d}A}{{\rm d}t}
  = \lambda_{1} A + \left( \widetilde{\Psi}, \mathcal{N}[f] \right)
\end{align}
and the equation for the unstable manifold $H$ as
\begin{align}
  \frac{{\rm d}H}{{\rm d}t}
  = \mathcal{L}H + \mathcal{N}[f]
  - \left[
    \left( \widetilde{\Psi}, \mathcal{N}[f] \right) \Psi
    + \left( \widetilde{\Psi}^{\ast}, \mathcal{N}[f] \right) \Psi^{\ast}
  \right].
  \label{eq:unstable-manifold-H}
\end{align}
These equations can be solved perturbatively for sufficiently small $|A|$,
and the right-hand-side of \eqref{eq:amplitude-equation-general}
is expanded as
\begin{align}
  \label{eq:amplitude-equation-complex}
  \frac{{\rm d}A}{{\rm d}t}
  = \lambda_{1} A + c_{3} A|A|^{2} + c_{5} A|A|^{4} + c_{7} A|A|^{6} + \cdots,
\end{align}
where the eigenvalue $\lambda_{1}$ and the coefficients $c_{3},c_{5},\cdots$
depend on the coupling constant $K$.
Note that the right-hand-side of \eqref{eq:amplitude-equation-complex}
has only odd order terms.
See Appendix \ref{sec:amp-eq} for derivations
and the explicit forms of coefficients.

The complex amplitude equation \eqref{eq:amplitude-equation-complex} can be
reduced to a real equation written as
\begin{align}
  \label{eq:amplitude-equation}
  \frac{{\rm d}\sigma}{{\rm d}t} = 2 \sigma G(\sigma),
\end{align}
where $\sigma=|A|^{2}\geq 0$ and
\begin{align}
  \label{eq:G-sigma}
  G(\sigma) = {\rm Re}(\lambda_{1}) + {\rm Re}(c_{3})\sigma
  + {\rm Re}(c_{5})\sigma^{2} + \cdots.
\end{align}
We will search for stationary solutions
with which the right-hand-side of the amplitude equation
\eqref{eq:amplitude-equation} is zero.
The nonsynchronized state,
for instance, corresponds to $\sigma=0$,
which always satisfies the stationary condition.

The amplitude equation is useful to determine
the continuity of the synchronization transition
from the nonsynchronized state $f^{0}$ 
to a partially synchronized state.
Around the critical point $K_{\rm c}$,
the order parameter must be small if the transition is continuous
and we can use the truncated equation
\begin{align}
  {\rm Re}(\lambda_{1}) + {\rm Re}(c_{3}) \sigma = 0.
\end{align}
  This equation has a nontrivial solution for ${\rm Re}(c_{3})<0$
  from the instability condition ${\rm Re}(\lambda_{1})>0$,
  while no real solution for ${\rm Re}(c_{3})>0$
  suggests a discontinuous transition.
Assuming continuity of $c_{3}(K)$ with respect to $K$
and taking the limit $K\to K_{\rm c}+0$, we say
that the synchronization transition is continuous
if ${\rm Re}(c_{3}(K_{\rm c}))<0$,
and is discontinuous if ${\rm Re}(c_{3}(K_{\rm c}))>0$ \cite{barre2016}.
  The separating point $(\gamma_{1},\Omega)=(1,1)$
between the domains A and B 
on the line $\gamma_{1}=1$
is obtained by searching for
the point which satisfies ${\rm Re}(c_{3}(K_{\rm c}))=0$.


\section{Criteria for determining domains}
\label{sec:criteria}

Looking back the five bifurcation diagrams exhibited in Fig.~\ref{fig:bif-diagrams},
we have two elements to characterize the diagrams,
which are oscillation and a jump of the order parameter.
We first discuss mechanisms of the oscillation and of the jump
in Secs.~\ref{sec:oscillation} and \ref{sec:jump} respectively
with the aid of the linear and nonlinear analyses 
reviewed in the previous section \ref{sec:analysis}.
After that, we propose a procedure to divide the parameter plane
into the five domains in Sec.~\ref{sec:division}.



\subsection{Oscillation of order parameter}
\label{sec:oscillation}

The asymptotically periodic oscillation of the order parameter
has been also observed in the Hamiltonian mean-field model,
and the oscillation comes from existence of two small clusters
running with different velocities \cite{morita-kaneko-06}.
Inspired by this work, we propose the following idea;
The first unstable eigenvalue induces a partially synchronized state
and the oscillation is excited by appearance of a second unstable eigenvalue.

The order parameter is computed as
\begin{align}
  z(t)
  = 2\pi \int_{-\infty}^{\infty} \widetilde{f}_{-1}(\omega,t) \diff\omega,
\end{align}
the time evolution of the order parameter $z(t)$ is hence described
by the eigenvalues arisen from the Fourier $-1$ mode.
As defined in the end of Sec.~\ref{eq:linear-eigenvalues},
the two fake eigenvalues, the roots of $D_{-1}(\lambda)$,
are denoted by $\lambda_{1}^{\ast}$ and $\lambda_{2}^{\ast}$,
where ${\rm Re}(\lambda_{1}^{\ast})\geq{\rm Re}(\lambda_{2}^{\ast})$.

Let us increase the value of the coupling constant $K$
from the nonsynchronized region $K<K_{\rm c}$.
Beyond the critical point $K_{\rm c}$,
the first fake eigenvalue $\lambda_{1}^{\ast}$ becomes unstable
and the resonant oscillators,
whose natural frequencies are close to ${\rm Im}(\lambda_{1}^{\ast})$,
form a cluster.
For instance, if $g(\omega)$ is symmetric and unimodal,
the resonant frequency is zero
and oscillators around $\omega=0$ form a synchronized cluster.
This small cluster corresponds to the continuous synchronization
transition in the bifurcation diagram (see Fig.~\ref{fig:bif-diagrams} (d)).
Further increasing $K$, for suitable pairs of $(\gamma_{1},\Omega)$,
the second fake eigenvalue $\lambda_{2}^{\ast}$ also becomes unstable
and forms the second cluster.
The two resonant frequencies differ in general,
${\rm Im}(\lambda_{1}^{\ast})\neq {\rm Im}(\lambda_{2}^{\ast})$,
and the order parameter oscillates;
The order parameter is large when the two clusters are close
on $\mathbb{S}^{1}$, and is small when the clusters are in the
antiphase positions each other.
The symmetry of $g(\omega)$, in particular, results
  in inducing simultaneous destabilization of the two fake
  eigenvalues $\lambda_{1}^{\ast}$ and $\lambda_{2}^{\ast}$ at $K_{\rm c}$
  and yielding the bifurcation diagram Fig.~\ref{fig:bif-diagrams} (c).

Summarizing, existence of two unstable eigenvalues,
which mean two pairs of the unstable eigenvalues
by counting the roots of $D_{1}(\lambda)$ in addition to
the roots of $D_{-1}(\lambda)$,
suggests the oscillation of the order parameter.
To support this mechanism,
we investigate $K$ dependence of the fake eigenvalues
$\lambda_{1}^{\ast}$ and $\lambda_{2}^{\ast}$
at points chosen from the domains C, D, and E
of the parameter plane $(\gamma_{1},\Omega)$.

\begin{figure}[htbp]
  \begin{center}
  \includegraphics[width=8cm]{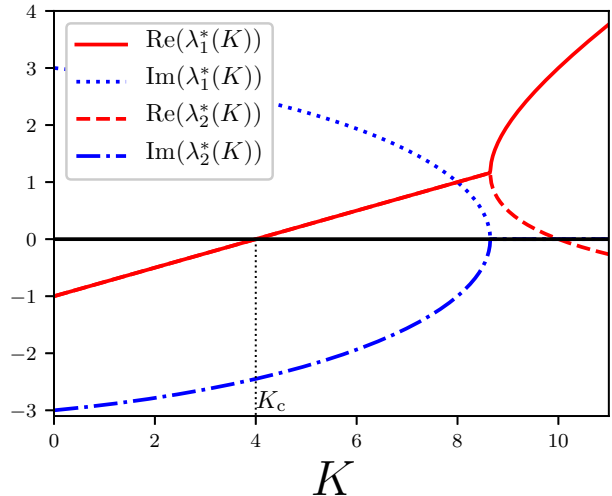}  
  \end{center}
  \caption{$K$ dependence of the
    fake eigenvalues $\lambda_{1}^{\ast}$ and $\lambda_{2}^{\ast}$
    for $(\gamma_{1},\Omega)=(1.0,3.0)$.
    The lines representing
    $\mathrm{Re}(\lambda_{1}^{\ast}(K))$ (red solid line)
    and $\mathrm{Re}(\lambda_{2}^{\ast}(K))$ (red dashed line)
    collapse for $K\leq 4(\sqrt{10}-1)\simeq 8.649$.
    The two fake eigenvalues $\lambda_{1}^{\ast}$ and $\lambda_{2}^{\ast}$
    become unstable simultaneously at $K_{\rm c}=4$.
  }
  \label{fig:ev1030}
\end{figure}

A symmetric case is examined in Fig.~\ref{fig:ev1030}
for $(\gamma_{1},\Omega)=(1.0,3.0)$ which belongs to the domain C.
As we expected, two eigenvalues become unstable
at the same $K_{\rm c}$ with different imaginary parts.
In Fig.~\ref{fig:phase-diagram}
the separating point $\Omega=\sqrt{3}$ between the domains B and C
on the symmetry line $\gamma_{1}=1$ is obtained
by checking existence of the two unstable eigenvalues.

\begin{figure}[htbp]
  \begin{center}  
  \includegraphics[width=8cm]{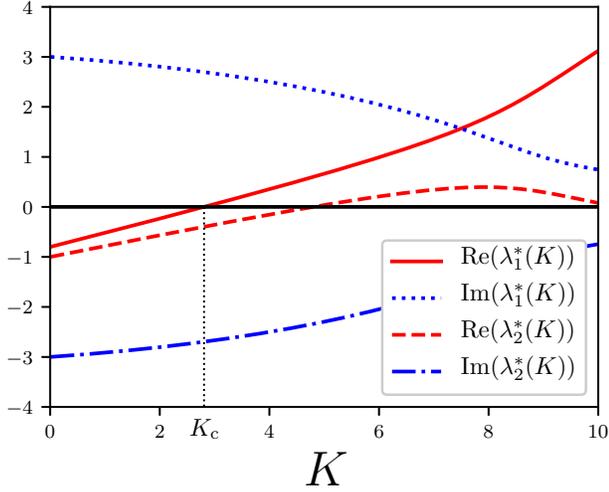}
  \end{center}
  \caption{$K$ dependence of the
    fake eigenvalues $\lambda_{1}^{\ast}$ and $\lambda_{2}^{\ast}$
    for $(\gamma_{1},\Omega)=(0.8,3.0)$ belonging to the domain D.
    The first fake eigenvalue $\lambda^{\ast}_{1}$ becomes unstable
    at the critical point $K_{\rm c}=2.807$,
    then the second fake eigenvalue $\lambda^{\ast}_{2}$ becomes
    unstable at a larger $K=4.794$.
  }
  \label{fig:ev0830}
\end{figure}

\begin{figure}[htbp]
\begin{center}
  \includegraphics[width=8cm]{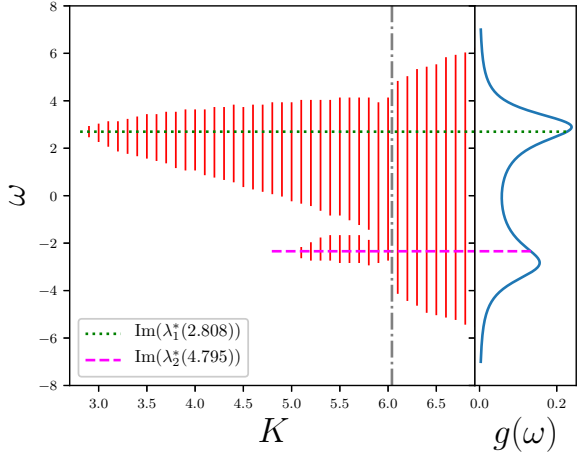}
\end{center}
  \caption{
    The natural frequency distribution and the emerging clusters
    for $(\gamma_{1},\Omega)=(0.8,3.0)$.
    The clusters, which are obtained by numerically integrating
    the $N$-body system \eqref{eq:kuramoto} with $N=10^{5}$,
    are found in the ranges of red bars.
    The second fake eigenvalue $\lambda_{2}^{\ast}$
    becomes unstable at $K=4.795$,
    which is approximately in agreement with the emergence point of the second lower cluster.
      Merging two clusters occurs around $K=6.04$, which is 
      plotted with gray dash-dotted line,
      and this leads to the jump in the order parameter,
      as seen in Fig.~\ref{fig:bif-diagrams}(d1).
  }
  \label{fig:cluster1}
\end{figure}

\begin{figure}[htbp]
\begin{center}
  \includegraphics[width=8cm]{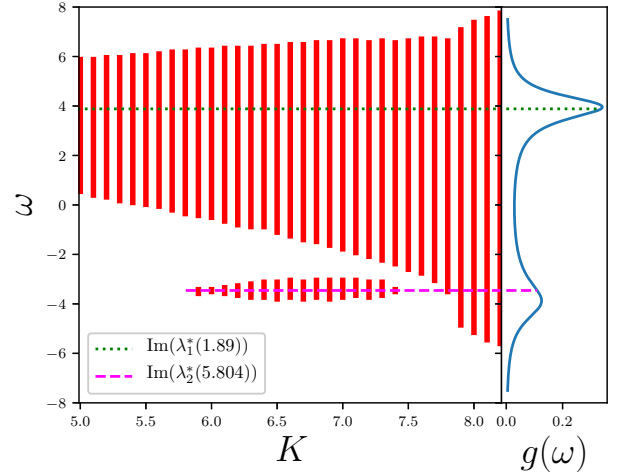}
\end{center}
  \caption{
    Same with Fig.~\ref{fig:cluster1} but for $(\gamma_{1},\Omega)=(0.6,4.0)$
    belonging to the domain D.
    The second fake eigenvalue $\lambda_{2}^{\ast}$
    becomes unstable at $K=5.804$,
    which is approximately in agreement with the emergence point of the second lower cluster.
    The second cluster does not merge
    into the first cluster,
    and the the second cluster vanishes around
    $K=7.5$.
    }
  \label{fig:cluster3}
\end{figure}

\begin{figure}[htbp]
\begin{center}
  \includegraphics[width=8cm]{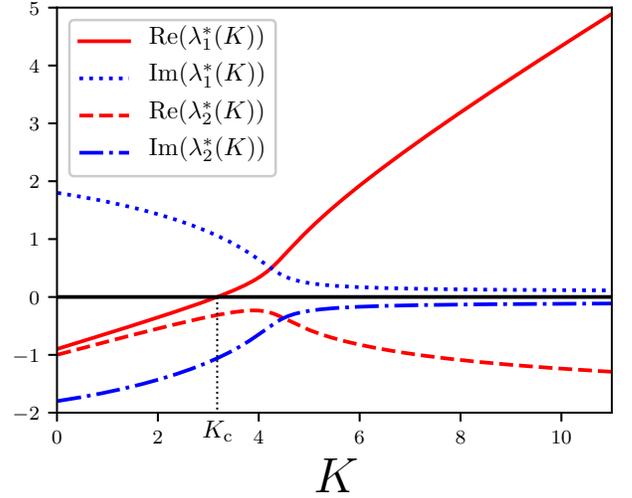}
\end{center}
  \caption{Same with Fig.~\ref{fig:ev0830}
    but for $(\gamma_{1},\Omega)=(0.9,1.8)$ belonging to the domain E.
    The real part of the fake eigenvalue $\lambda_{2}$ approaches to zero,
    but does not become positive.}
  \label{fig:ev0918}
\end{figure}

For the point $(\gamma_{1},\Omega)=(0.8,3.0)$
belonging to the domain D, 
the simultaneity of destabilization
breaks as shown in Fig.~\ref{fig:ev0830}
owing to asymmetry of $g(\omega)$.
A correspondence between the two unstable
eigenvalues and the two clusters is exhibited in Fig.~\ref{fig:cluster1},
where $N$-body simulations of the Kuramoto model \eqref{eq:kuramoto}
were performed by using the fourth-order Runge-Kutta algorithm
with the time step $\Delta t=0.01$.
See Appendix \ref{sec:capture_cluster}
  for the procedure to determine the synchronized intervals of $\omega$
  reported as red bars in Fig.~\ref{fig:cluster1}.
The second unstable eigenvalue and the second cluster emerge
  at almost same values of $K$,
  and this observation is consistent with the discussion above.
  The scenario holds even when the first cluster is further developed
  as shown in Fig.~\ref{fig:cluster3} for the point $(\gamma_{1},\Omega)=(0.6,4.0)$,
  although the linear analysis is performed around the
  nonsynchronized state.
  It is worth noting that the second fake eigenvalue approaches
  to zero but does not become unstable
  at the point $(\gamma_{1},\Omega)=(0.9,1.8)$
  close to the domain D but belonging to the domain E
  as reported in Fig.~\ref{fig:ev0918}.

\begin{figure}[htbp]
\begin{center}
  \includegraphics[width=8cm]{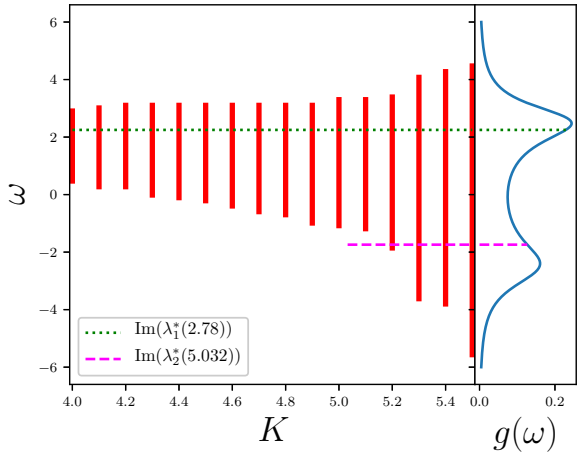}
\end{center}
  \caption{
    %
    Same with Fig.~\ref{fig:cluster1} but for $(\gamma_{1},\Omega)=(0.8,2.6)$
    belonging to the domain E.
    The second fake eigenvalue becomes unstable at $K=5.032$,
    but the second cluster does not appear.}
  \label{fig:cluster2}
\end{figure}

We note that the discussion above does not always hold
because the second unstable eigenvalue
does not always yield the second cluster.
At the point $(\gamma_{1},\Omega)=(0.8,2.6)$ belonging to the domain E,
at $K=5.032$, the second unstable eigenvalue emerges
but its imaginary part is not sufficiently far from
the grown first cluster,
and the second virtual cluster is absorbed by the first cluster
without emerging as shown in Fig.~\ref{fig:cluster2}.
One more condition must be therefore added to characterize the oscillation;
\begin{align}
  |{\rm Im}(\lambda_{1}^{\ast})-{\rm Im}(\lambda_{2}^{\ast})|>s
  \label{eq:imag-condition}
\end{align}
for some $s$ \cite{barre-yamaguchi-09}.
See Appendix~\ref{sec:imag-condition} for the further investigation on this condition.

\begin{figure}[htbp]
\begin{center}
  \includegraphics[width=8cm]{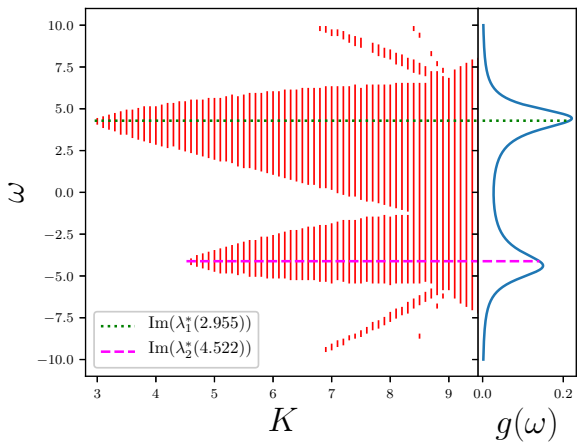}
\end{center}
  \caption{
    Same with Fig.~\ref{fig:cluster1} but for $(\gamma_{1},\Omega)=(0.82,4.5)$
    belonging to the domain E.
    The second fake eigenvalue $\lambda_{2}^{\ast}$
    becomes unstable at $K=4.522$,
    which is approximately in agreement with the emergence point of the second lower cluster.
    There exists clusters
    other than the first and second clusters corresponding
      to the two peaks of the natural frequency distribution $g(\omega)$.
    This multi-cluster state is referred to as the Bellerophon state \cite{bi2016,li2019}.}
  \label{fig:cluster5}
\end{figure}

  We give another remark on multi-cluster states;
  The maximum number of clusters is not necessarily two
  for the bimodal natural frequency distribution.
  Multi-cluster states called the Bellerophon states
  are reported in \cite{bi2016,li2019}
  for symmetric natural frequency distributions.
  The symmetry is, however, not essential to the multi-cluster states
  as an example is shown in Fig.~\ref{fig:cluster5}
  for an asymmetric distribution with $(\gamma_{1},\Omega)=(0.82,4.5)$.

\subsection{Jump of order parameter}
\label{sec:jump}

A jump of the order parameter emerges from $r=0$ (domain B)
or from $r>0$ (domain E) .
We expect that the jump from $r=0$ is identified
by ${\rm Re}(c_{3}(K_{\rm c}))>0$
as discussed in Sec.~\ref{sec:amplitude-equation}.
This criterion is successfully used in a generalized model \cite{barre2016},
but for symmetric natural frequency distributions.
We will verify this criterion for asymmetric ones with $\gamma<1$.

\begin{figure}[htbp]
\begin{center}
  \includegraphics[width=8cm]{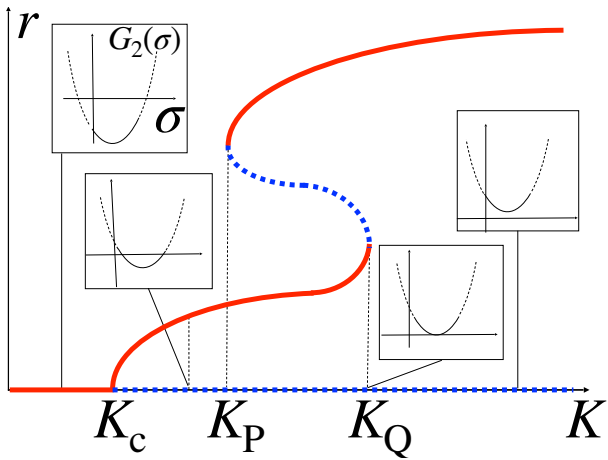}
\end{center}
\caption{
  Schematic picture for the bifurcation diagram of type E.
  The red solid lines denote the stable branches,
  and the blue dashed lines denote the unstable branches.
  Insets are schematic graphs of $G_{2}(\sigma)$.  
  }
  \label{fig:schematic2}
\end{figure}

To characterize the jump from $r>0$,
a typical bifurcation diagram of the type E is schematically shown
in Fig.~\ref{fig:schematic2}.
We focus on the saddle-node bifurcation point $K_{\rm Q}$,
at which $G(\sigma)$ in the amplitude equation \eqref{eq:amplitude-equation}
has degenerated nontrivial roots.
The degeneracy can be approximately captured by $G_{2}(\sigma)$,
the truncation of $G(\sigma)$ up to the second order of $\sigma$,
as the vanishing discriminant.

Let us discuss validity of the expansion \eqref{eq:amplitude-equation-complex}
around the saddle-node bifurcation point $K_{\rm Q}$.
The point $K_{\rm Q}$ is approximated by $K_{\rm Q'}$ obtained from $G_{2}(\sigma)$.
The value of $\sigma$ at $K_{\rm Q}$ is also approximated as
\begin{align}
  \sigma_{\ast}' = - \dfrac{{\rm Re}(c_{3}(K_{\rm Q'}))}{{\rm Re}(c_{5}(K_{\rm Q'}))}
\end{align}
and gives the equality
\begin{align}
  \label{eq:ratio-35}
  \dfrac{|{\rm Re}(c_{5}(K_{\rm Q'}))(\sigma_{\ast}')^{2}|}
  {|{\rm Re}(c_{3}(K_{\rm Q'}))\sigma_{\ast}'|} = \dfrac{1}{2}.
\end{align}
This equality suggests that the expansion \eqref{eq:amplitude-equation-complex}
up to the third term is not excellent but
does not break around the bifurcation point $K_{\rm Q}$.
We will discuss validity of \eqref{eq:amplitude-equation-complex} later
from the view point of smallness of the order parameter.

We remark on the upper stable branch appearing
at another saddle-node bifurcation point $K_{\rm P}$.
Coexistence of the three branches between $K_{\rm P}$ and $K_{\rm Q}$
is a distinguishing feature of the type E,
but we focus on $K_{\rm Q}$ because of the two advantages:
(i) Capturing the coexistence in the amplitude equation
\eqref{eq:amplitude-equation} requires the third order term of $G(\sigma)$,
the coefficient $c_{7}$, which is hard to compute.
(ii) The upper branch has a larger value of $\sigma$ than $\sigma_{\ast}'$,
and validity of the expansion \eqref{eq:amplitude-equation-complex}
becomes worse accordingly.

\subsection{Theoretical division of the parameter plane}
\label{sec:division}

In the basis of the discussions done in Secs.~\ref{sec:oscillation} and \ref{sec:jump},
we divide the parameter plane $(\gamma_{1},\Omega)$
into the five domains by the following flow.
For preparation, we compute the critical point $K_{\rm c}$
for a given pair of parameters $(\gamma_{1},\Omega)$
as shown in Appendix \ref{sec:Kc}.

We first focus on the critical point $K=K_{\rm c}$.
The discontinuous jump appears at $K_{\rm c}$ only in the type B
among the considered five bifurcation diagrams
(see Fig.~\ref{fig:bif-diagrams}).
The point $(\gamma_{1},\Omega)$ must belong to the domain B
if ${\rm Re}(c_{3}(K_{\rm c}))>0$ holds, which suggests a jump from $r=0$.
When this condition does not hold,
we check if there are two unstable eigenvalues at $K_{\rm c}$.
If this condition is satisfied, oscillation of the order parameter
starts from $K_{\rm c}$ and the point $(\gamma_{1},\Omega)$
is considered to be in the domain C.

When both checks at $K_{\rm c}$ are negative, 
we increase $K$ from $K_{\rm c}$ and examine the following two propositions:
\begin{align}
  \mathrm{(Oscillation) }\ 
  \exists K_{\rm O} (>K_{\rm c})\ 
  \mathrm{ s.t. }\ 
  {\rm Re}(\lambda_{2}(K_{\rm O}))>0,
  \label{eq:proposition-oscillation}
\end{align}
and
\begin{align}
  \mathrm{(Jump) }\ 
  \exists K_{\rm J} (>K_{\rm c})\ 
  \mathrm{ s.t. }\ 
  \Delta(K_{\rm J}) = 0,
  \label{eq:proposition-jump}
\end{align}
where the discriminant $\Delta(K)$ of $G_{2}(\sigma)$ is defined by
\begin{align}
  \Delta(K)
  = {\rm Re}(c_{3}(K))^{2} - 4{\rm Re}(\lambda(K)){\rm Re}(c_{5}(K)).
\end{align}
Note that the degenerated root of $G_{2}(\sigma)$, $-c_{3}/(2c_{5})$,
is positive under the assumptions $c_{3}(K_{\rm J})<0$ and $c_{5}(K_{\rm J})>0$.
If both the propositions \eqref{eq:proposition-oscillation}
and \eqref{eq:proposition-jump} are false, we decide that the point
$(\gamma_{1},\Omega)$ belongs to the domain A.
If the oscillation proposition \eqref{eq:proposition-oscillation}
is true but the jump proposition \eqref{eq:proposition-jump} is false,
the point must be in the domain D.
If the oscillation proposition \eqref{eq:proposition-oscillation}
is false but the jump proposition \eqref{eq:proposition-jump} is true,
the point must be in the domain E.
If both the propositions are true, there is a competition
between $K_{\rm O}$ and $K_{\rm J}$; $K_{\rm O}<K_{\rm J}$ suggests the domain D
and $K_{\rm O}>K_{\rm J}$ suggests the domain E.
  In Table~\ref{table:tech-domain}, 
  we summarize techniques which are used to determine the domains


\begin{table}
  \caption{Techniques we use to determine the domains.
  $\lambda_{1}$ corresponds to the linear theory, and
  $c_{3}$ and $c_{5}$ correspond to the nonlinear theory.
  }
  \label{table:tech-domain}
  \begin{tabular}{c||c|c|c|c}
   & $\lambda_{1}$ & $c_{3}$ & $c_{5}$ & Phenomenology \\\hline\hline
   \begin{minipage}{20mm}
    \scalebox{0.2}{\includegraphics{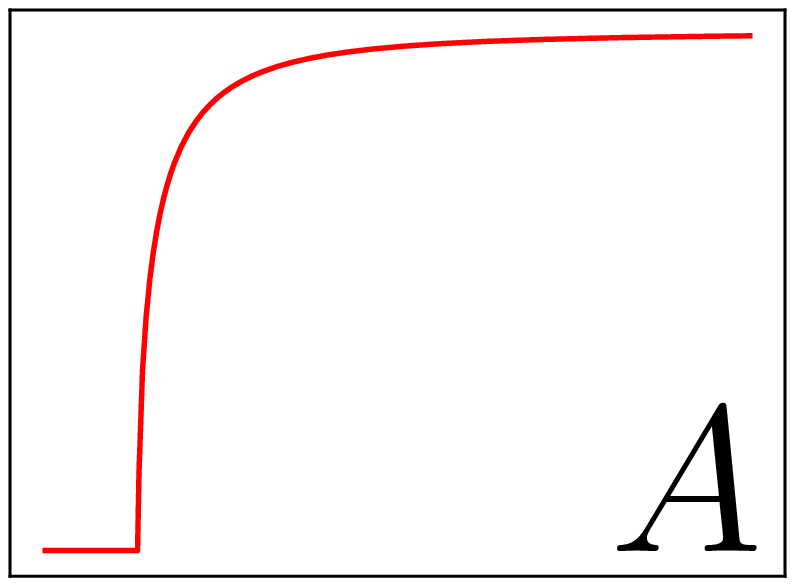}}
  \end{minipage}
  & \checkmark & \checkmark & \checkmark & \checkmark \\\hline
  \begin{minipage}{20mm}
    \scalebox{0.2}{\includegraphics{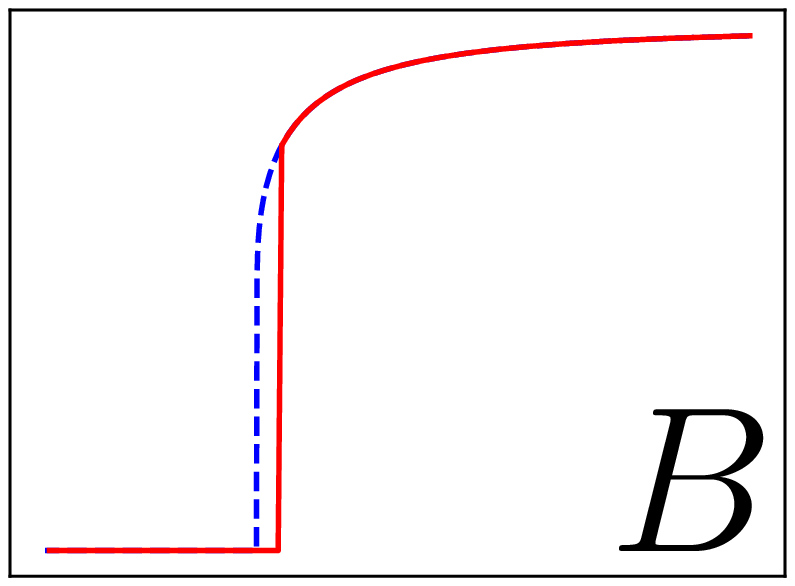}}
  \end{minipage}
   & \checkmark & & & \\\hline
   \begin{minipage}{20mm}
    \scalebox{0.2}{\includegraphics{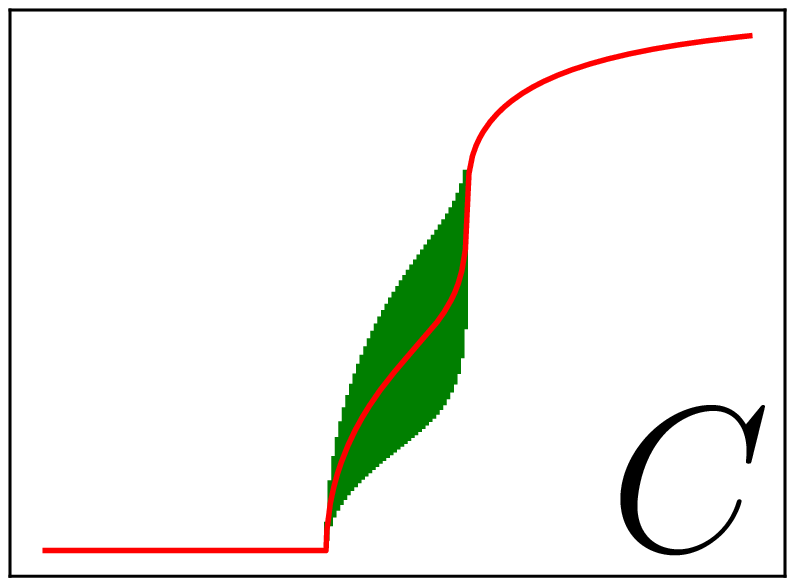}}
  \end{minipage}
   & \checkmark & & & \checkmark \\\hline
   \begin{minipage}{20mm}
    \scalebox{0.2}{\includegraphics{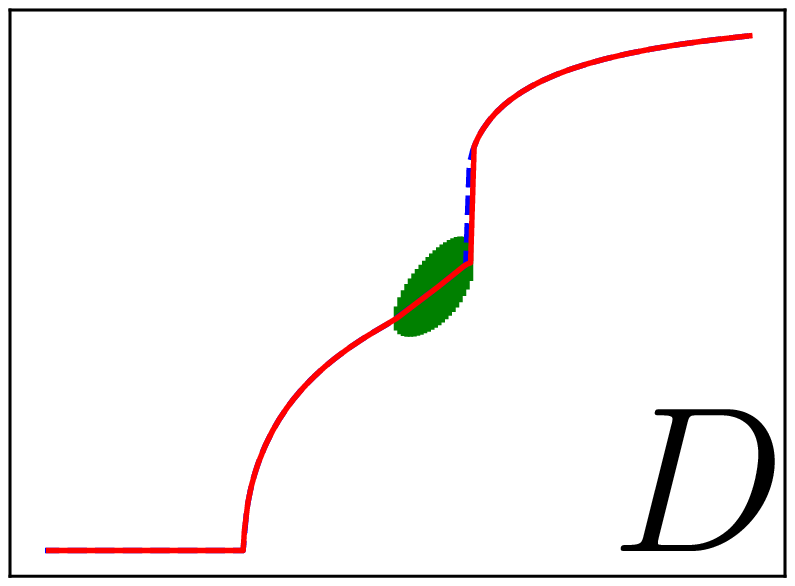}}
  \end{minipage}
   & \checkmark & \checkmark & \checkmark & \checkmark \\\hline
   \begin{minipage}{20mm}
    \scalebox{0.2}{\includegraphics{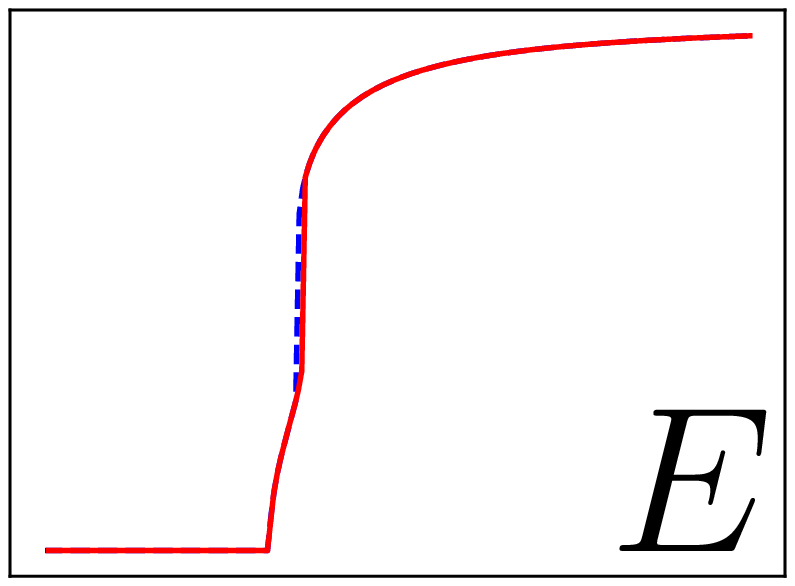}}
  \end{minipage}
   & \checkmark & \checkmark & \checkmark & \\\hline
  \end{tabular}
\end{table}

\begin{figure}[htbp]
\begin{center}
  \includegraphics[width=8cm]{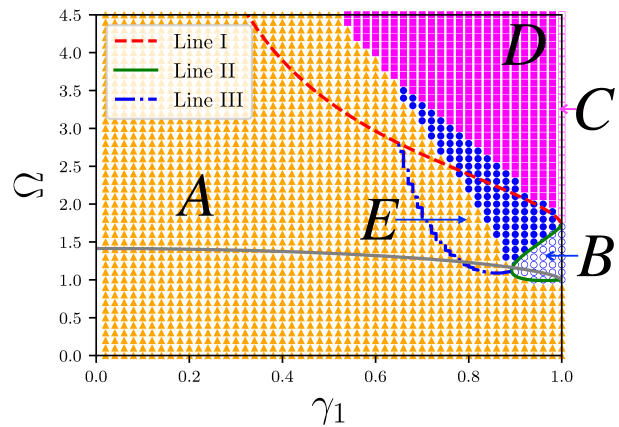}
\end{center}
\caption{
    The theoretical lines I(red broken line),
    II(green solid line), and III(blue dash-dotted line) are overlapped
    on Fig.~\ref{fig:phase-diagram}.
    Line I denotes the borderline between 
    the domains D and ${\rm A}\cup{\rm B}\cup{\rm E}$.
    Line II denotes the borderline between 
    the domains B and ${\rm A}\cup{\rm C}\cup{\rm E}$.
    Line III denotes the borderline between 
    the domains E and A.
    The gray line represents the borderline between the unimodal and bimodal natural frequency distributions $g(\omega)$.
  }
  \label{fig:pd-aa}
\end{figure}

The flow proposed above provides three theoretical lines,
I, II, and III
which divide the parameter plane $(\gamma_{1},\Omega)$ into the five domains,
as reported in Fig.~\ref{fig:pd-aa}.
  the upper side of the line I is the theoretical domain D,
  the inside of the line II is the theoretical domain B,
  and the lines I, II, and III enclose the theoretical domain E.
The theoretical lines reproduce qualitatively
the numerically obtained domains.

For the line I, our result is fairly consistent with numerical ones,
but the domain D is overestimated.
Our method says that
the emergence of two clusters is associated with 
the emergence of two unstable eigenvalues.
The second eigenvalue, however, does not always
  give rise to the second cluster as we discussed
  in Sec.~\ref{sec:oscillation}.
We have to reduce the domain D by introducing
an additional condition that
$|{\rm Im}(\lambda_{1}^{\ast})-{\rm Im}(\lambda_{2}^{\ast})|$
is sufficiently large.

The line II is perfect because the order parameter is small
around the critical point of the synchronization transition
and the perturbatively obtained amplitude equation is valid
for such a small amplitude.
This criterion had been used for symmetric $g(\omega)$,
but we have confirmed that it is also powerful for asymmetric ones.
The line II also reveals existence of discontinuous
synchronization transition for unimodal but asymmetric distributions.
The existence has been pointed out numerically in a previous study \cite{terada2018},
but our theoretical analysis ensures it.

\begin{figure}
\begin{center}
\includegraphics[width=8cm]{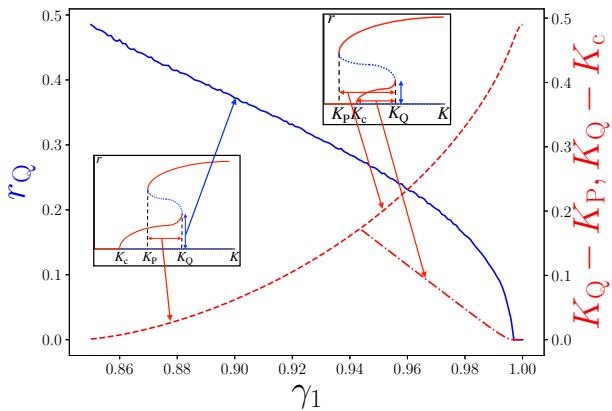}
\end{center}
\caption{The absolute value $r$ of the order parameter
  at $K_{\mathrm{Q}}$(blue solid line)
  and the value $K_{\mathrm{Q}}-K_{\mathrm{P}}$(red dashed line)
  for $\Omega=1.7$.
  As $\gamma_{1}$ decreases,
  $r_{\mathrm{Q}}$, the smaller value of $r$ at $K=K_{\mathrm{Q}}$,
  increases monotonically,
  and the value $K_{\mathrm{Q}}-K_{\mathrm{P}}$
  decreases monotonically.
  The red dash-dotted line represents the value
  $K_{\mathrm{Q}}-K_{\mathrm{c}}$,
  which is drawn only for $\gamma_{1}$ giving $K_{\mathrm{P}}<K_{\mathrm{c}}$;
  The upper stable branch emerges
  before the lower unstable branch emerges.
  }
  \label{fig:jump}
\end{figure}

The line III is not perfect.
We used a perturbative method to derive the amplitude equation
\eqref{eq:amplitude-equation-complex},
assuming that the order parameter is sufficiently small
at the jumping point $K_{\rm Q}$.
This assumption becomes worse as the point $(\gamma_{1},\Omega)$
approaches to the boundary of the domain E
where the width of the hysteresis vanishes
as reported in Fig.~\ref{fig:jump}.
It is thus expectable that the boundary of the domain E
is not perfectly described by the theoretical line III.
Nevertheless, the theoretical line III captures the domain E qualitatively
and is useful to have a strong candidate of the domain E.


\section{Conclusion and Discussions}
\label{sec:conclusion}

We have proposed a theoretical method to classify bifurcation diagrams
in the Kuramoto model
by using the amplitude equation with the aid of the linear analysis
around the nonsynchronized reference state.
The amplitude equation, obtained perturbatively,
is usually stopped up to the second leading term
because it is sufficient to judge the continuity
of the synchronization transition, bifurcation from the nonsynchronized state.
The premise, however, is broken by introducing asymmetry
  in the natural frequency distribution
  because the asymmetry induces bifurcations from partially synchronized states.
We have extended the amplitude equation up to the third leading term
and successfully captured a discontinuous bifurcation after a continuous one.

For bifurcation diagrams having oscillating states,
we have focused on unstable eigenvalues of the nonsynchronized state.
Roughly speaking, one unstable eigenvalue corresponds to one cluster formation,
and hence existence of two unstable eigenvalues suggest
appearance of two clusters rotating with different speeds.
This idea is qualitatively in good agreement with numerical results,
but overestimates the domain having the oscillation.
We have to reduce the domain by adding one more condition
which requires a large discrepancy of rotating speeds of the two clusters
in order to avoid the absorption of the second virtual cluster
by the first grown cluster.
The additional condition has to be investigated.
We remark that a similar condition was discussed in a Hamiltonian system
\cite{barre-yamaguchi-09}.

  We mainly used the amplitude equation to predict the bifurcation diagram.
  This method is a perturbation technique
  and is not powerful if the considering point is far from the critical point.
  This restriction prevents us from getting precise theoretical lines
  for $\Omega$ sufficiently large, for instance, because the oscillation/jump
  emerging from a partially synchronized state are not close to the critical point.
  On the contrary, for symmetric and bimodal natural frequency distributions,
  our method successfully identified the oscillating states,
  the domain C, solely by a linear analysis with phenomenology,
  while they have been previously analyzed by a nonlinear analysis,
  the amplitude equation \cite{crawford1994}.
  Our analysis hence captures an essence of the oscillating states.

It is worth noting that a cluster is created by
the resonance between natural frequencies an unstable eigenvalue,
but the number of clusters, $n_{c}$,
is not necessarily the same with the number of unstable eigenvalues, $n_{u}$.
In our computations, $n_{c}=1$ when $n_{u}=1$
but $n_{c}\geq 2$ when $n_{u}=2$.
Such a multi-cluster state with $n_{c}>2$, called the Bellerophon state, 
was reported previously \cite{bi2016,li2019}
but the mechanism is an open question.
Noting that two or more clusters induce oscillation of the order parameter
and following the scenario that a resonance makes a cluster,
we expect that the multi-cluster state with $n_{c}>2$
is realized hierarchically by resonances between natural frequencies
and the frequencies of the order parameter oscillation.
This scenario has to be investigated.

Finally, we discuss applications of our theory
to general coupled oscillator models.
Introducing the phase-lag parameter in the coupling function
\cite{sakaguchi-kuramoto-86},
we have a phase diagram having a continuous synchronization transition
followed by a discontinuous jump as the type E
\cite{omelchenko-wolfrum-12}.
Analyzing this system is a straightforward application of our theory.
Studying time delay \cite{yeung-strogatz-99,montbrio-pazo-schmidt-06}
is also an interesting application.
If the coupling function becomes general,
the coefficients of the amplitude equation may have
divergences as the coupling constant goes to the critical value
  as shown in a collisionless plasma \cite{crawford1995b}.
For instance, the coefficient ${\rm Re}(c_{3})$ may be proportional to
a power of $1/{\rm Re}(\lambda_{1})$.
  Nevertheless, the amplitude equation captures the asymptotic
  scaling of the order parameter against the strength of instability
  in plasmas, and we may expect that the amplitude equation still has power
  to predict the bifurcation diagram.
It is another future problem to check
the precision of the proposed criterion under such divergent coefficients.

\acknowledgments
Y.Y.Y. acknowledges the support of JSPS KAKENHI Grant No. 16K05472.

\appendix

\section{Ott-Antonsen ansatz and reduction}
\label{sec:oa}
  We derive the reduced system \eqref{eq:oa-12} from the equation of continuity \eqref{eq:kuramoto-inf} by using the Ott-Antonsen ansatz.
The Ott-Antonsen ansatz assumes the form of the probability distribution
$F(\theta,\omega,t)$ as
\begin{align}
  F = \frac{g(\omega)}{2\pi}
  \left\{ 1 + \sum_{k=1}^{\infty} \left[ a(\omega,t)^{k} e^{ik\theta}
      + a^{\ast}(\omega,t)^{k} e^{-ik\theta} \right] \right\}
  \label{eq:oa}
\end{align}
with $\left|a(\omega,t)\right|<1$ to ensure the convergence of the series.
By substituting (\ref{eq:oa}) to (\ref{eq:kuramoto-inf}), we have
\begin{align}
  \label{eq:eq-of-a}
  \frac{\partial a}{\partial t}+i\omega a+\frac{K}{2}\left(a^{2}z-z^{*}\right)=0,
\end{align}
where the order parameter is written as
\begin{align}
  z = \int_{-\infty}^{\infty}g(\omega)a^{\ast}(\omega,t) \diff\omega.
  \label{eq:oa-z}
\end{align}
The integration of the right-hand side can be computed
by using the residue theorem 
because $|g(\omega)a^{\ast}(\omega,t)|<g(\omega)$ holds
and $g(\omega)$ of \eqref{eq:g} decays faster than $1/|\omega|$
when $|\omega|\to\infty$.
The residue theorem gives
\begin{align}
  z(t)
  = k_{1} a^{\ast}(\Omega+i\gamma_{1},t)
  + k_{2} a^{\ast}(-\Omega+i\gamma_{2},t),
\end{align}
where the constants $k_{1}$ and $k_{2}$ are defined by \eqref{eq:k1k2}.
Introducing the complex variables $z_{1}$ and $z_{2}$ as
\begin{align}
  z_{1} = a^{\ast}(\Omega+i\gamma_{1},t),
  \quad
  z_{2} = a^{\ast}(-\Omega+i\gamma_{2},t),
\end{align}
the order parameter is expressed as
\begin{align}
  z = k_{1}z_{1}+k_{2}z_{2}.
\end{align}
Setting $\omega=\Omega+i\gamma_{1}$ and $\omega=-\Omega+i\gamma_{2}$
in \eqref{eq:eq-of-a}, we have dynamics of $z_{1}$ and $z_{2}$
as the equation \eqref{eq:oa-12}.

\section{Spectral functions and eigenfunctions}
\label{sec:spectral-functions}
  are obtained as roots of the spectral functions.
  We derive the spectral functions and the eigenfunction for an eigenvalue.
The linear operator $\mathcal{L}$ is expanded into the Fourier series as
\begin{align}
  \mathcal{L}f = \sum_{k\in\mathbb{Z}} \mathcal{L}_{k}\widetilde{f}_{k}(\omega,t) e^{ik\theta},
\end{align}
where $\widetilde{f}_{k}$ are the Fourier components of $f$ defined by
\eqref{eq:f-Fourier}.
The linear operator $\mathcal{L}_{k}$ is defined by
\begin{align}
  \label{eq:Lk}
  \mathcal{L}_{k} \widetilde{f}_{k}
  = -ik\omega \widetilde{f}_{k}
  + \frac{K}{2} g(\omega) ( \delta_{k,1} + \delta_{k,-1} )
  \int_{-\infty}^{\infty} \widetilde{f}_{k}(\omega,t) \diff\omega.
\end{align}
The symbol $\delta_{k,l}$ represents the Kronecker delta.

Once we have an eigenfunction $\psi(\omega)$ of the linear operator
$\mathcal{L}_{k}$ satisfying
\begin{align}
  \mathcal{L}_{k}\psi = \lambda \psi,
  \label{eq:eigenfunction-psi}
\end{align}
we have an eigenfunction $\Psi(\theta,\omega)$
of the linear operator $\mathcal{L}$ as
\begin{align}
  \Psi(\theta,\omega) = \psi(\omega)e^{ik\theta},
  \label{eq:Psi}
\end{align}
which induces
\begin{align}
  \mathcal{L}(\psi e^{ik\theta}) = \lambda (\psi e^{ik\theta}).
\end{align}
We therefore discuss eigenvalues and eigenfunctions
of the Fourier expanded linear operator $\mathcal{L}_{k}$.
The mode $k\neq\pm 1$ has $\mathcal{L}_{k}=-ik\omega$
and has only rotations. From now on, we focus on the modes $k=\pm 1$.


The equation \eqref{eq:eigenfunction-psi} is explicitly rewritten as
\begin{align}
  \label{eq:eigensystem}
  (\lambda + ik\omega) \psi(\omega)
  = \frac{K}{2} g(\omega) \int_{-\infty}^{\infty} \psi(\omega') \diff\omega'.
\end{align}
Let $\lambda+ik\omega\neq 0$.
Multiplying $(\lambda+ik\omega)^{-1}$ and integrating over $\omega$,
we have
\begin{align}
  \left( \int_{-\infty}^{\infty} \psi(\omega) \diff\omega \right)
  \Lambda_{k}(\lambda) = 0,
\end{align}
where the spectral functions $\Lambda_{\pm 1}(\omega)$
are defined in \eqref{eq:spectral-function}.
The vanishing integral,
$\int_{-\infty}^{\infty} \psi(\omega) ~\diff\omega=0$,
implies $\psi(\omega)\equiv 0$ for any $\omega\in\mathbb{R}$
from \eqref{eq:eigensystem} and the assumption $\lambda+ik\omega\neq 0$,
but this is not compatible with the assumption
that $\psi$ is an eigenfunction.
Consequently, the eigenvalue $\lambda$ must satisfy the equation 
$\Lambda_{k}(\lambda)=0$.
The nonvanishing integral may be assumed to be
$\int_{-\infty}^{\infty} \psi(\omega) ~\diff\omega=1$
without loss of generality,
and the eigenfunction $\psi$ is expressed as
\begin{align}
  \psi(\omega) = \frac{K}{2} \frac{g(\omega)}{\lambda+ik\omega},
  \quad
  (k=\pm 1).
  \label{eq:psi}
\end{align}

\section{Analytic continuation}
\label{sec:analytic-continuation}

We introduce the analytic continuation of the spectral functions
$\Lambda_{\pm 1}(\lambda)$, \eqref{eq:spectral-function}.
We start from a point $\lambda$ whose real part is positive,
${\rm Re}(\lambda)>0$, and decrease the real part.
The condition ${\rm Re}(\lambda)>0$ of the starting point
comes from the Laplace transform \cite{strogatz1992}
\begin{align}
  \widehat{f}(s) = \int_{0}^{\infty} f(t) e^{-st} \diff t,
\end{align}
which analyzes temporal evolution and is defined in the positive
real half-plane ${\rm Re}(s)>0$ to ensure convergence of the integral.
The variable $s$ corresponds to $\lambda$.

When $\lambda$ passes the imaginary axis,
the integrands of the spectral functions $\Lambda_{\pm 1}(\lambda)$,
\eqref{eq:spectral-function},
meet the singularities at $\omega=\pm i\lambda$.
To avoid this singularity at $\omega=i\lambda$
($\omega=-i\lambda$),
we modify continuously the integral contour
from the real axis by adding a small lower (upper) half-circle
around the singular point.
This modification brings a residue part to the integral.
Continuing this modification for ${\rm Re}(\lambda)<0$,
we have analytically continued functions of $\Lambda_{\pm 1}(\lambda)$ as
\begin{align}
  D_{\pm 1}(\lambda) = 1 - \frac{K}{2} I_{\pm 1}(\lambda),
  \label{eq:d_lambda}
\end{align}
where 
\begin{align}
  I_{\pm 1}(\lambda) = \left\{
    \begin{array}{ll}
      \displaystyle{\int_{-\infty}^{\infty} \frac{g(\omega)}{\lambda\pm i\omega} \diff\omega}, & {\rm Re}(\lambda)>0 \\
      \displaystyle{{\rm PV}\int_{-\infty}^{\infty} \frac{g(\omega)}{\lambda\pm i\omega} \diff\omega} + \pi g(\pm i\lambda) , & {\rm Re}(\lambda)=0 \\
      \displaystyle{\int_{-\infty}^{\infty} \frac{g(\omega)}{\lambda\pm i\omega} \diff\omega} + 2\pi g(\pm i\lambda) , & {\rm Re}(\lambda)<0 \\
    \end{array}
  \right.
  \label{eq:integral}
\end{align}
and PV represents the Cauchy principal value.
The roots of $D_{\pm 1}(\lambda)$ and of $\Lambda_{\pm 1}(\lambda)$
are identical from their definitions if ${\rm Re}(\lambda)>0$,
  and hence such roots of $D_{\pm 1}(\lambda)$ are eigenvalues.
They do not coincide, however, for ${\rm Re}(\lambda)\leq 0$ in general.
A root of $D_{\pm 1}(\lambda)$ with ${\rm Re}(\lambda)\leq 0$
is not an eigenvalue but is called
a resonance pole, a Landau pole, or a fake eigenvalue \cite{ogawa2013}.

The difference between $\Lambda_{1}(\lambda)$ and $D_{1}(\lambda)$
is demonstrated by considering a Lorentzian natural frequency distribution,
\begin{align}
  g(\omega) = \frac{\gamma}{\pi} \frac{1}{\omega^{2}+\gamma^{2}}.
\end{align}
Straightforward computations give
\begin{align}
  \Lambda_{1}(\lambda) = \left\{
    \begin{array}{ll}
      \displaystyle{1 - \frac{K}{2(\lambda+\gamma)}}, & {\rm Re}(\lambda)>0 \\
      \displaystyle{1 - \frac{K}{2(\lambda-\gamma)}}, & {\rm Re}(\lambda)<0 \\
    \end{array}
  \right.
\end{align}
and
\begin{align}
  D_{1}(\lambda) = 1 - \frac{K}{2(\lambda+\gamma)}, \quad \lambda\in\mathbb{C}.
\end{align}
$\lambda=K/2-\gamma$ is a root of $D_{1}(\lambda)$
but is not of $\Lambda_{1}(\lambda)$ if $K\leq K_{\rm c}$
and ${\rm Re}(\lambda)\leq 0$ accordingly,
where $K_{\rm c}=2\gamma$ is the synchronization transition point.
Similarly, we can reproduce the critical point $K_{\rm c}=2/[\pi g(0)]$
for a symmetric unimodal $g(\omega)$
by computing the roots of $D_{\pm 1}(\lambda)$.
The critical point $K_{\rm c}$ for the considering family \eqref{eq:g}
is given in Appendix \ref{sec:Kc}.


\section{Fake eigenvalues and synchronization transition point $K_{\rm c}$}
\label{sec:Kc}
  The fake eigenvalues are roots of the continued spectrum functions,
  \eqref{eq:d_lambda}. They move continuously on the complex plane,
  and the synchronization transition point $K_{\rm c}$ is computed
  as the point where one of the fake eigenvalue passes the imaginary axis
  from left to right with the smallest coupling constant $K>0$.
  The fake eigenvalues are analyzed in Appendix \ref{sec:fake-eigenvalues}
  for the natural frequency distribution \eqref{eq:g}.
  We will show that the first fake eigenvalue,
  which has the largest real part,
  passes the imaginary axis at most once time.
  After that, the equation to determine $K_{\rm c}$ will be derived
  in Appendix \ref{sec:synchroniztion-transition-point}
  with an analysis of the movement of the second fake eigenvalue.

\subsection{Fake eigenvalues}
\label{sec:fake-eigenvalues}

Substituting \eqref{eq:g} to \eqref{eq:integral},
we obtain the explicit form of $D_{1}(\lambda)$ as
\begin{align}
  D_{1}(\lambda)
  = 1 - \frac{K}{2\gamma^{+}}
  \frac{\gamma^{+}\lambda + (\gamma^{+})^{2}+i\Omega\gamma^{-}}
  {(\lambda+\gamma_{1}+i\Omega)(\lambda+\gamma_{2}-i\Omega)}
\end{align}
where
\begin{align}
  \gamma^{+} = \gamma_{1} + \gamma_{2} > 0,
  \quad
  \gamma^{-} = \gamma_{1} - \gamma_{2} \leq 0.
\end{align}
The equation $D_{1}(\lambda)=0$ induces the quadratic equation
\begin{align}
  \lambda^{2} - b(K)\lambda - a_{\rm R}(K) - ia_{\rm I}(K) = 0
  \label{eq:quadratic-equation-lambda}
\end{align}
where
\begin{align}
  \begin{aligned}
    & b(K) = \frac{K}{2} - \gamma^{+}, \\
    & a_{\rm R}(K) = \frac{K}{2} \gamma^{+} - (\gamma_{1}\gamma_{2}+\Omega^{2}), \\
    & a_{\rm I}(K) = \Omega\gamma^{-} \left( 1 + \frac{K}{2\gamma^{+}} \right) \leq 0.
  \end{aligned}
\end{align}

To write down the two solutions to \eqref{eq:quadratic-equation-lambda},
we introduce the complex variable
\begin{align}
  x = b^{2} + 4a_{\rm R} + i 4a_{\rm I} = \rho e^{i\theta},
  \quad
  \rho,\theta\in\mathbb{R}.
\end{align}
The argument $\theta$ is in the interval $[\pi,2\pi]$ from $a_{\rm I}\leq 0$.
The two fake eigenvalues $\lambda_{1}$ and $\lambda_{2}$,
which satisfy ${\rm Re}(\lambda_{1})\geq{\rm Re}(\lambda_{2})$,
are written as
\begin{align}
\begin{aligned}
    2\lambda_{1} = b - \sqrt{\rho} \cos\frac{\theta}{2}
    - i \sqrt{\rho} \sin\frac{\theta}{2}, \\
    2\lambda_{2} = b + \sqrt{\rho} \cos\frac{\theta}{2}
    + i \sqrt{\rho} \sin\frac{\theta}{2},
\end{aligned}
\end{align}
as $\cos(\theta/2)\leq 0$. 
Note that the signs of the imaginary parts are
${\rm Im}(\lambda_{1})\leq 0$ and ${\rm Im}(\lambda_{2})\geq 0$
from $\sin(\theta/2)\geq 0$
and that they are consistent with
Figs.~\ref{fig:ev1030}-\ref{fig:cluster2}.
We remark that the frequency of the order parameter
  corresponds to the imaginary part of not $\lambda_{1}$ but $\lambda_{1}^{\ast}$.

The synchronization transition point $K_{\rm c}$ is determined
by the equation ${\rm Re}(\lambda_{1}(K_{\rm c}))=0$.
Let us show that the solution is at most one.
Using the relation
\begin{align}
  \cos\frac{\theta}{2} = - \sqrt{\frac{1+\cos\theta}{2}}
\end{align}
and the definitions of $\rho$ and $\theta$,
\begin{align}
  \rho = \sqrt{(b^{2}+4a_{\rm R})^{2}+(4a_{\rm I})^{2}},
  \quad
  \cos\theta = \frac{b^{2}+4a_{\rm R}}{\rho},
\end{align}  
we have
\begin{align}
  2{\rm Re}(\lambda_{1}(K))
  = b + \frac{1}{\sqrt{2}}
  \sqrt{ \sqrt{(b^{2}+4a_{\rm R})^{2}+(4a_{\rm I})^{2}} + b^{2}+4a_{\rm R} }.
\end{align}
The functions $b(K),$ $a_{\rm I}^{2}(K)$, and $b^{2}(K)+4a_{\rm R}(K)$
are increasing functions of $K$ for $K>0$ from
\begin{align}
  b^{2}+4a_{\rm R}
  = \left( \frac{K}{2} + \gamma^{+} \right)^{2} - 4(\gamma_{1}\gamma_{2}+\Omega^{2}).
\end{align}
These increasing functions imply that
the real part ${\rm Re}(\lambda_{1}(K))$ is
also an increasing function of $K$ for $K>0$
and takes zero at most once time.

\subsection{Synchronization transition point $K_{\rm c}$}
\label{sec:synchroniztion-transition-point}
We show that there exists the unique solution to
the equation ${\rm Re}(\lambda_{1}(K))=0$,
which determines the synchronization transition point $K_{\rm c}$.
At $K_{\rm c}$, the fake eigenvalue $\lambda$ must be
pure imaginary of the form
$\lambda=i\lambda_{\rm I}~(\lambda_{\rm I}\in\mathbb{R})$.
Substituting this form with $K=K_{\rm c}$ into the quadratic equation
\eqref{eq:quadratic-equation-lambda}, we have
\begin{align}
  -\lambda_{\rm I}^{2} - ib(K_{\rm c})\lambda_{\rm I}
  - a_{\rm R}(K_{\rm c}) - i a_{\rm I}(K_{\rm c}) = 0.
\end{align}
The real part reads
\begin{align}
  \lambda_{\rm I}^{2} + a_{\rm R}(K_{\rm c}) = 0
  \label{eq:lambda-critical-real}
\end{align}
and the imaginary part reads
\begin{align}
  b(K_{\rm c})\lambda_{\rm I} + a_{\rm I}(K_{\rm c}) = 0.
  \label{eq:lambda-critical-imag}
\end{align}
Eliminating $\lambda_{\rm I}$ from 
\eqref{eq:lambda-critical-real} and \eqref{eq:lambda-critical-imag},
we have the cubic equation to determine $K_{\rm c}$ as
\begin{align}
  \label{eq:kc-poly}
  \begin{aligned}
    & (\gamma^{+})^{3} K_{\rm c}^{3}
    - 2 \left\{ 2(\gamma^{+})^{4} + \gamma_{1}\gamma_{2}
      \left[ (\gamma^{+})^{2} + 4\Omega^{2} \right] \right\} K_{\rm c}^{2}\\
    & + 4\gamma^{+} \left[
      (\gamma^{+})^{4} + 2\gamma_{1}\gamma_{2}(\gamma^{+})^{2}
      + 4\Omega^{2}\left( \gamma_{1}^{2}+\gamma_{2}^{2} \right) \right] K_{\rm c} \\
    & -8\gamma_{1}\gamma_{2}(\gamma^{+})^{2} \left[
      (\gamma^{+})^{2}+4\Omega^{2} \right] = 0.
  \end{aligned}
\end{align}
The number of real solutions to \eqref{eq:kc-poly} is one or three,
and all the real solutions are positive
because the left-hand side of \eqref{eq:kc-poly}
is always negative for $K_{\rm c}<0$.
The synchronization transition point is determined
as the smallest real solution to \eqref{eq:kc-poly}.

To investigate the number of unstable eigenvalues,
we introduce the discriminant $\Delta_{3}$ for the cubic equation
\eqref{eq:kc-poly}. In general, the discriminant is defined as
\begin{align}
  \Delta_{3} = b^{2}c^{2} - 27a^{2}d^{2} - 4ac^{3} - 4b^{3}d + 18abcd
\end{align}
for the cubic equation
\begin{align}
  ax^{3}+bx^{2}+cx+d=0.
\end{align}
The number of real solutions is one for $\Delta_{3}<0$
and is three for $\Delta_{3}>0$.

In the case $\Delta_{3}<0$ the first fake eigenvalue $\lambda_{1}$
passes the imaginary axis at the unique real solution $K_{\rm c}$
and no other passing occurs.
In the case $\Delta_{3}>0$ we have three different real solutions
of $K_{\rm c}^{(1)}$, $K_{\rm c}^{(2)}$, and $K_{\rm c}^{(2)}$,
where $K_{\rm c}^{(1)}<K_{\rm c}^{(2)}<K_{\rm c}^{(3)}$.
By the definition, the first fake eigenvalue $\lambda_{1}$
passes the imaginary axis at $K_{\rm c}=K_{\rm c}^{(1)}$
and it does not pass the imaginary axis any more
as shown in the previous subsection \ref{sec:fake-eigenvalues}.
As a result, the other two solutions $K_{\rm c}^{(2)}$ and $K_{\rm c}^{(3)}$
are realized by the second fake eigenvalue $\lambda_{2}$;
It passes the imaginary axis from left to right at $K_{\rm c}^{(2)}$,
and from right to left at $K_{\rm c}^{(3)}$,
as observed in Fig.~\ref{fig:ev0830}.

\section{Adjoint linear operator}
\label{sec:adjoint}
  The idea of deriving the amplitude equation presented
  in Sec.~\ref{sec:amplitude-equation} is to project the full dynamics
  onto the reduced space spanned by some eigenfunctions of
  the linear operator $\mathcal{L}$, \eqref{eq:linear-L}.
  The projection operator is defined by eigenfunctions
  of the adjoint operator of $\mathcal{L}$.
  In this Appendix, a necessary eigenfunction of the adjoint operator
  $\mathcal{L}^{\dagger}$ is presented.

The adjoint operator $\mathcal{L}^{\dagger}$ acts on $f$ as
\begin{align}
  \mathcal{L}^{\dagger}f
  = \omega\frac{\partial f}{\partial\theta}
  + \frac{K}{2}\left(r_{1}[ff^{0}]e^{-i\theta}+r_{-1}[ff^{0}]e^{i\theta}\right),
\end{align}
where
\begin{align}
  r_{n}[f] = \int_{-\infty}^{\infty} \diff\omega \int_{-\pi}^{\pi} \diff\theta~
  f(\theta,\omega,t) e^{in\theta}.
\end{align}
As done for the linear operator $\mathcal{L}$,
we expand $\mathcal{L}^{\dagger}$ into the Fourier series as
\begin{align}
  \mathcal{L}^{\dagger}f
  = \sum_{k\in\mathbb{Z}}
  \mathcal{L}^{\dagger}_{k}\widetilde{f}_{k}(\omega,t) e^{ik\theta}.
\end{align}
The linear operator $\mathcal{L}^{\dagger}_{k}$ is defined by
\begin{align}
  \mathcal{L}^{\dagger}_{k} \widetilde{f}_{k}
  = ik\omega \widetilde{f}_{k}
  + \frac{K}{2} \left[ \delta_{k,1} + \delta_{k,-1} \right]
  \int_{-\infty}^{\infty} \widetilde{f}_{k}(\omega,t) g(\omega) \diff\omega.
\end{align}
We focus on the modes $k=\pm 1$.

Let $\mu$ be an eigenvalue of $\mathcal{L}_{k}^{\dagger}$
and $\widetilde{\psi}(\omega)$ be the corresponding eigenfunction
which satisfies
\begin{align}
  \mathcal{L}_{k}^{\dagger}\widetilde{\psi} = \mu \widetilde{\psi}.
\end{align}
This equation brings
\begin{align}
  (\mu-ik\omega) \widetilde{\psi}(\omega)
  = \frac{K}{2} \int_{-\infty}^{\infty}
  \widetilde{\psi}(\omega) g(\omega) \diff\omega.
  \label{eq:tildepsi-equation}
\end{align}
Repeating the same discussion done in Appendix \ref{sec:spectral-functions},
the eigenvalue $\mu$ must satisfy
\begin{align}
  \Lambda_{k}^{\ast}(\mu^{\ast}) = 0.
\end{align}
This equation implies that $\mu^{\ast}$ is an eigenvalue
of $\mathcal{L}^{\dagger}_{k}$ if $\mu$ is an eigenvalue of $\mathcal{L}_{k}$.

Let $\lambda$ be an eigenvalue of $\mathcal{L}_{1}$
and $\psi$ be the corresponding eigenfunction \eqref{eq:psi}.
From the above discussion,
$\mathcal{L}_{1}^{\dagger}$ has an eigenvalue $\lambda^{\ast}$
and the corresponding eigenfunction is computed as
\begin{align}
  \widetilde{\psi}(\omega)
  = \frac{1}{[\Lambda'(\lambda)]^{\ast}} \frac{1}{\lambda^{\ast}-i\omega}.
  \label{eq:tildepsi}
\end{align}
As the case of $\mathcal{L}$,
$\lambda^{\ast}$ is also an eigenvalue of $\mathcal{L}^{\dagger}$
and the corresponding eigenfunction is
\begin{align}
  \widetilde{\Psi}(\theta,\omega)
  = \frac{\widetilde{\psi}(\omega)}{2\pi} e^{i\theta},
\end{align}
which satisfies the normalization condition
\begin{align}
  \left( \widetilde{\Psi}, \Psi \right) = 1.
  \label{eq:normalization-condition}
\end{align}


\section{Derivation of the amplitude equation}
\label{sec:amp-eq}

  The coefficients of the amplitude equation \eqref{eq:amplitude-equation-complex}
  are obtained perturbatively with an expression of the unstable manifold.
  We give explicit forms of the coefficients.

\subsection{Derivations of equations for $A$ and $H$}

We assume that $\lambda$ is the unique unstable eigenvalue of $\mathcal{L}_{1}$
and $\psi(\omega)$ is the corresponding eigenfunction.
The relation $\Lambda_{-1}(\lambda^{\ast})=\Lambda_{1}^{\ast}(\lambda)$
implies that $\lambda^{\ast}$ is an eigenvalue of $\mathcal{L}_{-1}$
and $\psi^{\ast}(\omega)$ is the corresponding eigenfunction.
The linear operator $\mathcal{L}$, therefore, has two unstable eigenvalues
$\lambda$ and $\lambda^{\ast}$, and the corresponding eigenfunctions respectively
\begin{align}
  \Psi(\theta,\omega) = \psi(\omega) e^{i\theta},
  \quad
  \Psi^{\ast}(\theta,\omega) = \psi^{\ast}(\omega) e^{-i\theta}.
\end{align}
Using these eigenfunctions, we expand the perturbation $f$
into the form of \eqref{eq:expand-f},
\begin{align}
  f = A(t)\Psi + A^{\ast}(t)\Psi^{\ast} + H(\theta,\omega,A,A^{\ast}).
\end{align}
We assume that the unstable manifold $H$ is tangent
to the unstable eigenspace ${\rm Span}(\Psi,\Psi^{\ast})$
at $A=A^{\ast}=0$.
Substituting this expansion into the equation of continuity, we have
\begin{align}
    \frac{{\rm d}A}{{\rm d}t} \Psi
    + \frac{{\rm d}A^{\ast}}{{\rm d}t} \Psi^{\ast}
    + \frac{{\rm d}H}{{\rm d}t} 
    & = \lambda A \Psi + \lambda^{\ast}A^{\ast}\Psi^{\ast}
    + \mathcal{L}H + \mathcal{N}[f].
  \label{eq:expanded-eq-continuity}
\end{align}

In \eqref{eq:expanded-eq-continuity},
taking the inner product with $\widetilde{\Psi}$,
we obtain the equation for $A$ as
\begin{align}
  \frac{{\rm d}A}{{\rm d}t}
  = \lambda A + \left( \widetilde{\Psi}, \mathcal{N}[f] \right).
  \label{eq:dAdt-appendix}
\end{align}
Extracting this equality and its complex conjugate
from \eqref{eq:expanded-eq-continuity},
we have the equation for $H$ as
\begin{align}
  \frac{{\rm d}H}{{\rm d}t}
  = \mathcal{L}H + \mathcal{N}[f]
  - \left[ \left( \widetilde{\Psi}, \mathcal{N}[f] \right) \Psi
    + \left( \widetilde{\Psi}^{\ast}, \mathcal{N}[f] \right) \Psi^{\ast}
  \right].
  \label{eq:dHdt-appendix}
\end{align}
The left-hand side of \eqref{eq:dHdt-appendix} is read as
\begin{align}
  \frac{{\rm d}H}{{\rm d}t}
  = \frac{\partial H}{\partial A} \frac{{\rm d}A}{{\rm d}t}
  + \frac{\partial H}{\partial A^{\ast}} \frac{{\rm d}A^{\ast}}{{\rm d}t}.
\end{align}
The right-hand side of \eqref{eq:dAdt-appendix} 
starts from the linear term $\lambda A$,
while one of \eqref{eq:dHdt-appendix} starts from the quadratic terms
of $A$ and $A^{\ast}$ by the tangency assumption of the unstable manifold,
$H=O(|A|^{2})$.
We, therefore, solve the two equations
\eqref{eq:dAdt-appendix} and \eqref{eq:dHdt-appendix}
perturbatively by assuming that $|A|$ is small.

\subsection{Fourier series expansion of $H$}
Before going to the Taylor series expansion with respect to $|A|$,
we expand $H$ into the Fourier series as
\begin{align}
  H = \sum_{n\in\mathbb{Z}} H_{n}(\omega,A,A^{\ast}) e^{in\theta}.
\end{align}
The rotational symmetry of the system yields
\cite{crawford1994}
\begin{align}
  H_{n}(\omega,A,A^{*})=
  \begin{cases}
    0 & n=0\\
    A\sigma h_{1}(\sigma,\omega) & n=1\\
    A^{n}h_{n}(\sigma,\omega) & n\geq 2,
  \end{cases}
\end{align}
where $\sigma=|A|^{2}$.

\begin{figure}
  \centering
  \begin{picture}(240,70)
    \put(30,60){$c_{3}$}
    \put(30,30){$c_{5}$}
    \put(30,0){$c_{7}$}
    \put(60,-5){\line(0,1){70}}
    \put(120,60){\underline{$h_{2,0}$}}
    \put(120,50){\vector(-2,-1){20}}
    \put(125,50){\vector(0,-1){10}}
    \put(80,30){\underline{$h_{1,0}$}}
    \put(120,30){\underline{$h_{2,1}$}}
    \put(160,30){$h_{3,0}$}
    \put(100,32){\vector(1,0){15}}
    \put(155,32){\vector(-1,0){15}}
    \put(85,20){\vector(0,-1){10}}
    \put(125,20){\vector(0,-1){10}}
    \put(165,20){\vector(0,-1){10}}
    \put(120,20){\vector(-2,-1){20}}
    \put(100,20){\vector(2,-1){20}}
    \put(160,20){\vector(-2,-1){20}}
    \put(140,20){\vector(2,-1){20}}
    \put(80,0){\underline{$h_{1,1}$}}
    \put(120,0){\underline{$h_{2,2}$}}
    \put(160,0){$h_{3,1}$}
    \put(200,0){$h_{4,0}$}
    \put(100,2){\vector(1,0){15}}
    \put(155,2){\vector(-1,0){15}}
    \put(195,2){\vector(-1,0){15}}
  \end{picture}
  \caption{Hierarchy of $h_{n,m}$.
    Each line represents $h_{n,m}$ which newly appear
    to compute the coefficient written in the most left column.
    Arrows indicate dependency, although some arrows are omitted
    for a graphical reason.
    The functions with underlines are directly used in \eqref{eq:c3c5c7}.}
  \label{fig:hierarchy-of-hmn}
\end{figure}
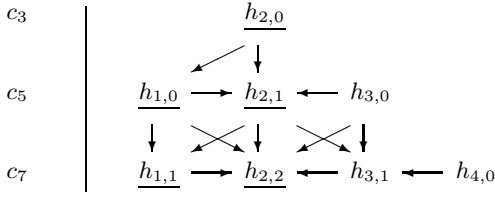

Now we expand $h_{n}$ into the Taylor series.
For sufficiently small $\sigma$ we expand 
\begin{align}
  h_{n}(\sigma,\omega)
  = h_{n,0}(\omega) + \sigma h_{n,1}(\omega) + \cdots.
  \label{eq:Taylor-expansion-hn}
\end{align}
Algebraic computations gives the equation of $A$ as
\begin{align}
  \frac{{\rm d}A}{{\rm d}t}
  = \lambda A + c_{3} A \sigma + c_{5} A \sigma^{2} + c_{7} A \sigma^{3} + \cdots
\end{align}
where
\begin{align}
\begin{aligned}
  c_{3}=&-\pi K\langle\widetilde{\psi},h_{2,0}\rangle,\\
  c_{5}=&-\pi K\left[\langle\widetilde{\psi},h_{2,0}\rangle\left(\int h_{1,0}\diff\omega\right)^{*}+\langle\widetilde{\psi},h_{2,1}\rangle\right],\\
  c_{7}=&-\pi K\left[\left(\int h_{1,1} \diff\omega\right)^{*}\langle\widetilde\psi,h_{2,0}\rangle+\langle\widetilde\psi,h_{2,2}\rangle\right.\\
  &+\left.\left(\int h_{1,0} \diff\omega\right)^{*}\langle\widetilde\psi,h_{2,1}\rangle\right],
  \label{eq:c3c5c7}
\end{aligned}
\end{align}
and
\begin{align}
  \langle f_{1}, f_{2} \rangle
  = \int_{-\infty}^{\infty} f_{1}^{\ast} f_{2} \diff\omega.
\end{align}
For simplicity of notation, we introduce a symbol
\begin{equation}
  \aveave{f} = \int_{-\infty}^{\infty} f \diff\omega.
\end{equation}
We compute
$\ave{ \widetilde{\psi}, h_{2,0}},~
\aveave{h_{1,0}}, ~
\ave{ \widetilde{\psi}, h_{2,1}}, ~
\aveave{h_{1,1}}$
and $\ave{ \widetilde\psi, h_{2,2}}$
by expanding \eqref{eq:dHdt-appendix} into the Fourier series.
Four Fourier modes 
  of $n=1, 2, 3$ and $4$
are sufficient to compute $c_{3}$, $c_{5}$, and $c_{7}$
  as shown in Fig.~\ref{fig:hierarchy-of-hmn}.


\subsection{Functions $h_{n,m}$}
The leading order of the Fourier mode $n$, $h_{n,0}(\omega)$, is expressed as
\begin{equation}
  \label{eq:hn0}
  h_{n,0}(\omega) = \dfrac{\pi^{n-1}K^{n}}{2} \dfrac{g(\omega)}{(\lambda+i\omega)^{n}},
  \qquad (2\leq n\leq 4).
\end{equation}
The other $5$ functions appearing in Fig.~\ref{fig:hierarchy-of-hmn}
solve the following equations:
\begin{equation}
  (2\lambda + \lambda^{\ast}+i\omega) h_{1,0}
  = \dfrac{K}{2} g(\omega) \aveave{h_{1,0}} - \pi K h_{2,0} - c_{3} \psi,
\end{equation}
\begin{equation}
  \begin{split}
    & (3\lambda + \lambda^{\ast}+2i\omega) h_{2,1} \\
    & = 2\pi K \left( \aveave{h_{1,0}} \psi + h_{1,0} - h_{3,0} \right)
    - 2c_{3} h_{2,0},
  \end{split}
\end{equation}
\begin{equation}
  \begin{split}
    & (3\lambda + 2\lambda^{\ast}+i\omega) h_{1,1}
    = \dfrac{K}{2} g(\omega) \aveave{h_{1,1}} \\
    & - \pi K \left( \aveave{h_{1,0}}^{\ast} h_{2,0} + h_{2,1} \right)
    - (2c_{3}+c_{3}^{\ast}) h_{1,0} - c_{5} \psi
  \end{split}
\end{equation}
\begin{equation}
  \begin{split}
    & 2(2\lambda + \lambda^{\ast}+i\omega) h_{2,2}
    = - (3c_{3}+c_{3}^{\ast}) h_{2,1} - 2c_{5} h_{2,0} \\
    & + 2\pi K \left( \aveave{h_{1,1}} \psi + h_{1,1} + \aveave{h_{1,0}} h_{1,0}
      - h_{3,1} - \aveave{h_{1,0}}^{\ast} h_{3,0}  \right),
  \end{split}
\end{equation}
and
\begin{equation}
  \begin{split}
    & (4\lambda + \lambda^{\ast}+3i\omega) h_{3,1} \\
    & = - 3c_{3} h_{3,0} + 3\pi K \left( \aveave{h_{1,0}} h_{2,0} + h_{2,1} - h_{4,0} \right).
  \end{split}
\end{equation}

\subsection{Inner products and integrals in the coefficients}
The expression \eqref{eq:hn0} and the eigenfunction $\widetilde{\psi}$,
\eqref{eq:tildepsi}, give the inner product
$\langle\widetilde{\psi},h_{2,0}\rangle$ as
\begin{equation}
  \langle\widetilde{\psi},h_{2,0}\rangle
  = - \dfrac{\pi K}{2} \dfrac{D_{1}''(\lambda)}{D_{1}(\lambda)}.
\end{equation}
To express the other quantities, we introduce two integral operators of
\begin{equation}
  \mathcal{I}_{m}[f](\lambda) = \int_{-\infty}^{\infty} \dfrac{f}{m\lambda+(m-1)\lambda^{\ast}+i\omega} \diff\omega
\end{equation}
and
\begin{equation}
  \mathcal{J}_{m}[f](\lambda) = \int_{-\infty}^{\infty} \dfrac{f}{(\lambda+i\omega)[m\lambda+\lambda^{\ast}+(m-1)i\omega]} \diff\omega.
\end{equation}
Remarking that $\aveave{h_{1,0}}$ and $\aveave{h_{1,1}}$ are solved
self-consistently and using the relation
\begin{equation}
  1 - \dfrac{K}{2} \mathcal{I}_{m}[g](\lambda)
  =  D_{1}(m\lambda+(m-1)\lambda^{\ast}),
\end{equation}
we have the necessary quantities as
\begin{equation}
  \aveave{h_{1,0}} = - \dfrac{\pi K \mathcal{I}_{2}[h_{2,0}] + c_{3}\mathcal{I}_{2}[\psi]}{D_{1}(2\lambda+\lambda^{\ast})},
\end{equation}
\begin{equation}
  \begin{split}
    & \ave{\widetilde{\psi}, h_{2,1}}
    = - \dfrac{2c_{3}}{D_{1}'(\lambda)} \mathcal{J}_{3}[h_{2,0}] \\
    & + \dfrac{2\pi K}{D_{1}'(\lambda)} \left(
      \aveave{h_{1,0}} \mathcal{J}_{3}[\psi]
      + \mathcal{J}_{3}[h_{1,0}]
      - \mathcal{J}_{3}[h_{3,0}] \right),
  \end{split}
\end{equation}
\begin{equation}
  \begin{split}
    & \aveave{h_{1,1}}
    = - \dfrac{(2c_{3}+c_{3}^{\ast}) \mathcal{I}_{3}[h_{1,0}] + c_{5} \mathcal{I}_{3}[\psi]}{D_{1}(3\lambda+2\lambda^{\ast})} \\
    & - \dfrac{\pi K}{D_{1}(3\lambda+2\lambda^{\ast})}
    \left( \aveave{h_{1,0}}^{\ast} \mathcal{I}_{3}[h_{2,0}]
      + \mathcal{I}_{3}[h_{2,1}] \right) \\
  \end{split}
\end{equation}
and
\begin{equation}
  \begin{split}
    & \ave{\widetilde{\psi},h_{2,2}}
    = - \dfrac{3c_{3}+c_{3}^{\ast}}{2D_{1}'(\lambda)} \mathcal{J}_{2}[h_{2,1}]
    - \dfrac{c_{5}}{D_{1}'(\lambda)} \mathcal{J}_{2}[h_{2,0}] \\
   & + \dfrac{\pi K}{D_{1}'(\lambda)} \left(
     \aveave{h_{1,1}} \mathcal{J}_{2}[\psi] + \mathcal{J}_{2}[h_{1,1}]
     + \aveave{h_{1,0}} \mathcal{J}_{2}[h_{1,0}] \right. \\
   & \hspace*{5em} \left.
     - \mathcal{J}_{2}[h_{3,1}]
     - \aveave{h_{1,0}}^{\ast} \mathcal{J}_{2}[h_{3,0}]
   \right),
  \end{split}
\end{equation}
where all the integral operators,
$\mathcal{I}_{m}$ and $\mathcal{J}_{m}$, are
evaluated at $\lambda$.
\section{Procedure to determine the synchronized intervals of $\omega$}
\label{sec:capture_cluster}

This Appendix summarizes the procedure to determine the intervals
of $\omega$, which we call the synchronized intervals,
in which oscillators synchronize.
The synchronized intervals are identified
through dividing the $\omega$-axis into bins of the width $w$.
The bin $\alpha$ defined by $[\alpha w, (\alpha+1)w)$
induces the set $P_{\alpha}$ consisting of the indices
of the oscillators included in this bin as
\begin{align}
  P_{\alpha} = \{ i~|~ i\in\{1,\cdots,N\},~
  \omega_{i} \in [\alpha w, (\alpha+1) w) \}.
\end{align}
We set $w=0.1$ here.

The oscillators in a bin gather around a point on the $(\theta,\omega)$ plane,
when the bin is in the synchronized intervals.
In the Kuramoto model, the gathering point with $\omega=(\alpha+1/2)w$
is unique for the bin $\alpha$,
and the standard deviation of the phases $\{\theta_{i}\}_{i\in P_{\alpha}}$,
denoted by $s_{\alpha}$, must be sufficiently small.
In contrast, the standard deviation $s_{\alpha}$ must be large
when the bin is not in the synchronized intervals.
We note that the standard deviation $s_{\alpha}$ is defined by
\begin{align}
  s_{\alpha}
  = \sqrt{ \dfrac{1}{|P_{\alpha}|} \sum_{i\in P_{\alpha}}
    (\theta_{i}-\mu_{\alpha})^{2} },
\end{align}
where $|P_{\alpha}|$ is the number of elements of the set $P_{\alpha}$
and $\mu_{\alpha}$ is the mean in the bin $\alpha$ as
\begin{align}
  \mu_{\alpha} = \dfrac{1}{|P_{\alpha}|} \sum_{i\in P_{\alpha}} \theta_{i}.
\end{align}
The standard deviation and the mean are computed
after taking mod $2\pi$ for each $\theta_{i}$.

We have to be careful for picking up all the bins
belonging to the synchronized intervals
because the standard deviation $s_{\alpha}$ becomes large
if the gathering point is close to $0$ or $2\pi$.
To overcome this difficulty, we prepare the shifted phases
\begin{align}
  \widetilde{\theta}_{i} = \theta_{i} + \pi \quad
  \text{(mod $2\pi$)}
\end{align}
and compute the standard deviation $\widetilde{s}_{\alpha}$ for
$\{\widetilde{\theta}_{i}\}_{i\in P_{\alpha}}$.
By using $s_{\alpha}$ and $\widetilde{s}_{\alpha}$,
we conclude that the bin $\alpha$ is in the synchronized intervals if
\begin{align}
  \min\{ s_{\alpha}, \widetilde{s}_{\alpha} \} \leq s_{\ast}
\end{align}
holds with $s_{\ast}=0.1$.

\section{Criterion of the imaginary difference (\ref{eq:imag-condition})}
\label{sec:imag-condition}

  The domain D is characterized by oscillation of the order parameter.
  As discussed in Sec.~\ref{sec:oscillation},
  the oscillation must have one more condition to satisfy,
  the imaginary difference \eqref{eq:imag-condition},
  to ensure emergence of the second cluster without being absorbed
  by the first cluster.
  We will show a trial to introduce this criterion
  to shave the theoretically obtained domain D
  (see Fig.~\ref{fig:pd-aa}).
  For the Kuramoto model with a unimodal and symmetric
  natural frequency distribution,
  only one unstable eigenvalue for $K>K_{\mathrm{c}}$ exists,
  resulting in only one cluster and the continuous transition.
  The half width of the cluster along the $\omega$ axis
  is calculated as $Kr$~\cite{strogatz2000}.
  Inspired by 
  this estimation, we introduce the difference criterion as
  \begin{align}
    |{\rm Im}(\lambda_{1}^{\ast})-{\rm Im}(\lambda_{2}^{\ast})|>K_{2}r(K_{2})
    \label{eq:imag-condition-1}
  \end{align}
  where $K_{2}$ is the emerging point of the second unstable eigenvalue.
  We estimate $r(K_{2})$, the value of order parameter at $K=K_{2}$,
  from the solution of the amplitude equation up to the third order, that is
  \begin{align}
    r(K_{2})=2\pi\sqrt{-\frac{{\rm Re}(\lambda_{1}(K_{2}))}{{\rm Re}(c_{3}(K_{2}))}}.
  \end{align}

\begin{figure}
  \includegraphics[width=8cm]{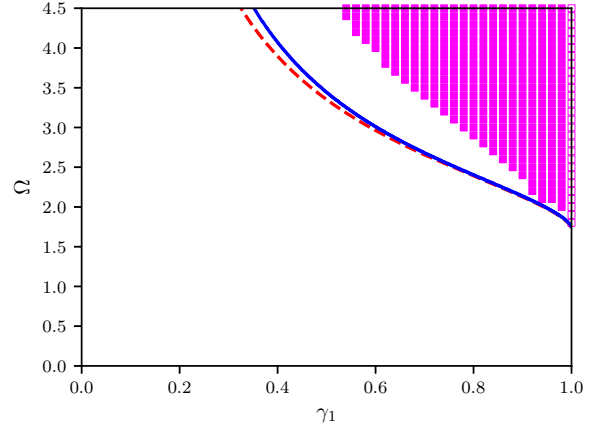}
  \caption{
      Comparison among a restricted domain D by the criterion
      \eqref{eq:imag-condition-1} (blue solid line),
      the original theoretical domain D (red broken line,
      which is the same with the line I reported in Fig.~\ref{fig:pd-aa}),
      and numerically obtained domain D (magenta squares),
      together with domain C (open magenta squares).
  }
  \label{fig:imag-condition}
\end{figure}

  In Fig.~\ref{fig:imag-condition},
  a domain D restricted by the criterion \eqref{eq:imag-condition-1}
  is compared with the original theoretical domain D.
  The criterion shaves a restricted area of the domain D,
  which is close to the original theoretical line I.
  The criterion still needs to be investigated.

\begin{thebibliography}{99}
\bibitem{pantaleone2002}
  J. Pantaleone, Synchronization of metronomes, \textit{Am. J. Phys.} \textbf{70}, 992 (2002).
\bibitem{smith1935}
  H. M. Smith, Synchronous flashing of fireflies, \textit{Science} \textbf{82}, 151 (1935).
\bibitem{buck1968}
  J. Buck and E. Buck, Mechanism of rhythmic synchronous flashing of fireflies, \textit{Science} \textbf{159}, 1319 (1968).
\bibitem{aihara2014}
  I. Aihara, T. Mizumoto, T. Otsuka, H. Awano, K. Nagira, H. G. Okuno and K. Aihara,
  Spatio-Temporal Dynamics in Collective Frog Choruses Examined by Mathematical Modeling and Field Observations,
  \textit{Sci. Rep.} {\bf 4}, 3891 (2014).
\bibitem{wiesenfeld1996}
  K. Wiesenfeld, P. Colet and S. H. Strogatz,
  Synchronization Transitions in a Disordered Josephson Series Array,
  \textit{Phys. Rev. Lett.} {\bf 76}, 404 (1996).
\bibitem{wiesenfeld1998}
  K. Wiesenfeld, P. Colet and S. H. Strogatz,
  Frequency locking in Josephson arrays: Connection with the Kuramoto model,
  \textit{Phys. Rev. E} {\bf 57}, 1563 (1998).
\bibitem{winfree1967}
  A. T. Winfree,
  Biological Rhythms and the Behavior of Populations of Coupled Oscillators,
  \textit{J. Theoret. Biol.} {\bf 16}, 15 (1967).
\bibitem{kuramoto1975}
  Y. Kuramoto, Self-entertainment of a population of coupled non-linear oscillators, \textit{International Symposium on Mathematical Problems in Theoretical Physics, Lecture Notes in Physics} \textbf{39}, 420 (1975).
\bibitem{strogatz2000}
  S. H. Strogatz, From Kuramoto to Crawford: exploring the onset of synchronization in populations of coupled oscillators, \textit{Physica} \textbf{143D}, 1 (2000).
\bibitem{chiba2015}
  H. Chiba, A proof of the Kuramoto conjecture for a bifurcation structure of the infinite-dimensional Kuramoto model, \textit{Ergod. Theo. Dyn. Syst.} \textbf{35}, 762 (2015).

\bibitem{basnarkov-urumov-07}
  L. Basnarkov and V. Urumov,
  Phase transitions in the Kuramoto model,
  \textit{Phys. Rev. E} \textbf{76}, 057201 (2007).
  
\bibitem{bastian-18}
  B. Pietras, N. Deschle, and A. Daffertshofer,
  First-order phase transitions in the Kuramoto model with compact bimodal frequency distributions,
  \textit{Phys. Rev. E} \textbf{98}, 062219 (2018).
  


\bibitem{martens2009}
  E. A. Martens, E. Barreto, S. H. Strogatz, E. Ott, P. So, and T. M. Antonsen, Exact results for the Kuramoto model with a bimodal frequency distribution, \textit{Phys. Rev. E} \textbf{79}, 026204 (2009).

\bibitem{terada2017}
  Y. Terada, K. Ito, T. Aoyagi, and Y. Y. Yamaguchi, Nonstandard transition in the Kuramoto model: a role of asymmetry in natural frequency distributions, \textit{J. Stat. Mech.} (2017) 0134403.

\bibitem{pazo-05}
  D. Paz\'{o},
  Thermodynamic limit of the first-order phase transition in the Kuramoto model,
  \textit{Phys. Rev. E} {\bf 72}, 046211 (2005).

\bibitem{park-kahng-19}
  J. Park and B. Kahng,
  Synchronization transitions through metastable state on structured networks,
  arXiv:1901.02123.

\bibitem{komarov-pikovsky-13}
  M. Komarov and A. Pikovsky,
  Multiplicity of Singular Synchronous States in the Kuramoto Model of Coupled Oscillators,
  \textit{Phys. Rev. Lett.} {\bf 111}, 204101 (2013).

\bibitem{komarov-pikovsky-14}
  M. Komarov and A. Pikovsky,
  The Kuramoto model of coupled oscillators with a bi-harmonic coupling function,
  \textit{Physica D} {\bf 289}, 18 (2014).

\bibitem{oa2008}
  E. Ott and T. M. Antonsen, Low dimensional behavior of large systems of globally coupled oscillators, \textit{Chaos} \textbf{18}, 037113 (2008).

\bibitem{oa2009}
  E. Ott and T. M. Antonsen,
  Long time evolution of phase oscillator systems,
  \textit{Chaos} {\bf 19}, 023117 (2009).

\bibitem{lu-etal-16}
  Z. Lu, K. Klein-Carde\~{n}a, S. Lee, T. M. Antonsen, M. Girvan, and E. Ott,
  Resynchronization of circadian oscillators and the east-west asymmetry of jet-lag,
  \textit{Chaos} \textbf{26}, 094811 (2016).
  
\bibitem{akao-19}
  A. Akao, S. Shirasaka, Y. Jimbo, B. Ermentrout, and K. Kotani,
  Theta-gamma cross-frequency coupling enables covariance between distant brain regions,
  arXiv:1903.12155.

\bibitem{crawford1994}
  J. D. Crawford, Amplitude Expansions for Instabilities in Populations of Globally-Coupled Oscillators, 
  \textit{J. Stat. Phys.} 
  \textbf{74}, 1047 (1994). 

\bibitem{crawford1995}
  J. D. Crawford,
  Scaling and Singularities in the Entrainment of Globally Coupled Oscillators,
  {\it Phys. Rev. Lett.} {\bf 74}, 4341 (1995).

\bibitem{metivier-19}
  D. M\'{e}tivier and S. Gupta,
  Bifurcations in the Time-Delayed Kuramoto Model of Coupled Oscillators: Exact Results,
  \textit{J. Stat. Phys.}, (2019).
  
  
\bibitem{barre2016}
  J. Barr\'{e} and D. M\'{e}tivier, Bifurcations and Singularities for Coupled Oscillators with Inertia and Frustration, \textit{Phys. Rev. Lett.} \textbf{117}, 214102 (2016).

\bibitem{crawford1994b}
  J. D. Crawford,
  Universal Trapping Scaling on the Unstable Manifold for a Collisionless Electrostatic Mode,
  {\it Phys. Rev. Lett.} {\bf 73}, 656 (1994).

\bibitem{crawford1995b}
  J. D. Crawford,
  Amplitude equations for electrostatic waves: Universal singular behavior in the limit of weak instability,
  {\it Phys. Plasmas} {\bf 2}, 97 (1995).

\bibitem{barre-metivier-yamaguchi-16}
  J. Barr{\'e}, D. M{\'e}tivier, and Y. Y. Yamaguchi,
  Trapping scaling for bifurcations in the Vlasov systems,
  \textit{Phys. Rev. E} {\bf 93}, 042207 (2016).


\bibitem{hansel-mato-meunier-93}
  D. Hansel, G. Mato and C. Meunier,
  Phase Dynamics for Weakly Coupled Hodgkin-Huxley Neurons,
  \textit{Europhys. Lett.} {\bf 23}, 367 (1993).

\bibitem{kiss-zhai-hudson-05}
  I. Z. Kiss, Y. Zhai and J. L. Hudson,
  Predicting Mutual Entrainment of Oscillators with Experiment-Based Phase Models,
  \textit{Phys. Rev. Lett.} {\bf 94}, 248301 (2005).

\bibitem{kiss-zhai-hudson-06}
  I. Z. Kiss, Y. Zhai and J. L. Hudson,
  Characteristics of Cluster Formation in a Population of Globally Coupled Electrochemical Oscillators: An Experiment-Based Phase Model Approach,
  \textit{Prog. Theor. Phys. Suppl.} {\bf 161}, 99 (2006).



\bibitem{terada2018}
  Y. Terada, K. Ito, R. Yoneda, T. Aoyagi and Y. Y. Yamaguchi,
  A role of asymmetry in linear response of globally coupled oscillator systems,
  arXiv:1802.08383.

\bibitem{lancellotti2004}
  C. Lancellotti, On the Vlasov Limit for Systems of Nonlinearly Coupled Oscillators without Noise, \textit{Transp. Theory Stat.} \textbf{34}, 523 (2004).


\bibitem{morita-kaneko-06}
  H. Morita and K. Kaneko,
  Collective Oscillation in a Hamiltonian System,
  \textit{Phys. Rev. Lett.} \textbf{96}, 050602 (2006).
  

  
  





  
\bibitem{barre-yamaguchi-09}
  J. Barr{\'e} and Y. Y. Yamaguchi,
  Small traveling clusters in attractive and repulsive Hamiltonian mean-field models,
  \textit{Phys. Rev. E} {\bf 79}, 036208 (2009).

\bibitem{bi2016}
H. Bi, X. Hu, S. Boccaletti, X. Wang,
Y. Zou, Z. Liu, and S. Guan,
Coexistence of Quantized, Time Dependent, Clusters in Globally Coupled Oscillators,
\textit{Phys. Rev. Lett.} \textbf{117}, 204101 (2016).

\bibitem{li2019}
X. Li, T. Qiu, S. Boccaletti,
I. Sendi-Nadal, Z. Liu and S. Guan,
Synchronization clusters emerge as the result of a global coupling among classical phase oscillators,
\textit{New J. Phys.} \textbf{21}, 053002 (2019).

\bibitem{sakaguchi-kuramoto-86}
  H. Sakaguchi and Y. Kuramoto,
  A Soluble Active Rotator Model Showing Phase Transitions via Mutual Entrainment,
  \textit{Prog. Theor. Phys.} {\bf 76}, 576 (1986).

\bibitem{omelchenko-wolfrum-12}
  O. E. Omel'chenko and M. Wlfrum,
  Nonuniversal Transitions to Synchrony in the Sakaguchi-Kuramoto Model,
  \textit{Phys. Rev. Lett.} {\bf 109}, 164101 (2012).

\bibitem{yeung-strogatz-99}
  M. K. S. Yeung and S. H. Strogatz,
  Time Delay in the Kuramoto Model of Coupled Oscillators,
  \textit{Phys. Rev. Lett.} {\bf 82}, 648 (1999).

\bibitem{montbrio-pazo-schmidt-06}
  E. Montbri{\'o}, D. Paz{\'o} and J. Schmidt,
  Time delay in the Kuramoto model with bimodal frequency distribution,
  \textit{Phys. Rev. E} {\bf 74}, 056201 (2006).

\bibitem{strogatz1992}
  S. H. Strogatz, R. E. Mirollo and P. C. Matthews,
  Coupled Nonlinear Oscillators below the Synchronization Threshold: Relaxation be Generalized Landau Damping,
  \textit{Phys. Rev. Lett.} {\bf 68}, 2730 (1992).

\bibitem{ogawa2013}
  S. Ogawa,
  Spectral and formal stability criteria of spatially inhomogeneous stationary solutions to the Vlasov equation for the Hamiltonian mean-field model,
  \textit{Phys. Rev. E} {\bf 87}, 062107 (2013). 

\end{thebibliography}
\end{document}